\documentclass[twocolumn]{pasj01}

\Received{$\langle$reception date$\rangle$}
\Accepted{$\langle$acception date$\rangle$}
\Published{$\langle$publication date$\rangle$}

\begin{document}

\title{A Molecular Line Survey toward the Nearby Galaxies NGC 1068, NGC 253, and IC 342 at 3 mm with the Nobeyama 45-m Radio Telescope:\\
 Impact of an AGN on 1 kpc Scale Molecular Abundances}
\author{%
  Taku \textsc{Nakajima},\altaffilmark{1,*}
  Shuro \textsc{Takano},\altaffilmark{2,3,$\dagger$}
  Kotaro \textsc{Kohno},\altaffilmark{4,5}
  Nanase \textsc{Harada},\altaffilmark{6}
  and
  Eric \textsc{Herbst}\altaffilmark{7}
}
\altaffiltext{1}{Institute for Space-Earth Environmental Research, Nagoya University, Furo-cho, Chikusa-ku, Nagoya, Aichi 464-8601, Japan}
\altaffiltext{2}{Nobeyama Radio Observatory, National Astronomical Observatory of Japan, 462-2, Nobeyama, Minamimaki, Minamisaku, Nagano 384-1305, Japan}
\altaffiltext{3}{Department of Astronomical Science, The Graduate University for Advanced Studies (SOKENDAI), 462-2, Nobeyama, Minamimaki, Minamisaku, Nagano 384-1305, Japan}
\altaffiltext{4}{Institute of Astronomy, The University of Tokyo, \\2-21-1, Osawa, Mitaka, Tokyo 181-0015, Japan}
\altaffiltext{5}{Research Center for Early Universe, School of Science, The University of Tokyo, \\7-3-1, Hongo, Bunkyo-ku, Tokyo 113-0033, Japan}
\altaffiltext{6}{Academia Sinica Institute of Astronomy and Astrophysics, P.O. Box 23-141, Taipei 10617, Taiwan}
\altaffiltext{7}{Department of Chemistry, University of Virginia, McCormick Road, P.O. Box 400319, \\Charlottesville, VA 22904, USA}
\email{nakajima@isee.nagoya-u.ac.jp}
\altaffiltext{$\dagger$}{Present address: Department of Physics, General Studies, College of Engineering, Nihon University, Tamuramachi, Koriyama, Fukushima 963-8642, Japan}


\maketitle

\begin{abstract}
It is important to investigate the relationships between the power sources and the chemical compositions of galaxies for understanding the scenario of galaxy evolution. We carried out an unbiased molecular line survey towards AGN host galaxy NGC1068, and prototypical starburst galaxies, NGC 253 and IC 342, with the Nobeyama 45-m telescope in the 3-mm band. The advantage of this line survey is that the obtained spectra have the highest angular resolution ever obtained with single-dish telescopes. In particular, the beam size of this telescope is $\sim$15$^{\prime\prime}$--19$^{\prime\prime}$, which is able to spatially separate the nuclear molecular emission from that of the starburst ring ($d$$\sim$30$^{\prime\prime}$) in NGC 1068. We successfully detected approximately 23 molecular species in each galaxy, and calculated rotation temperatures and column densities. We estimate the molecular fractional abundances with respect to $^{13}$CO and CS molecules and compare them among three galaxies in order to investigate the chemical signatures of an AGN environment. As a result, we found clear trends on the abundances of molecules surrounding the AGN on 1 kpc scale. HCN, H$^{13}$CN, CN, $^{13}$CN, and HC$_{3}$N are more abundant, and CH$_{3}$CCH is deficient in NGC 1068 compared with the starburst galaxies. High abundances of HCN, H$^{13}$CN, and HC$_{3}$N suggest that the circumnuclear disk in NGC 1068 is in a high-temperature environment. The reason for the non-detection of CH$_{3}$CCH is likely to be dissociation by high energy radiation or less sublimation of a precursor of CH$_{3}$CCH from grains.
\end{abstract}

\section{Introduction}

An unbiased line survey is an effective method in understanding chemical compositions as well as excitation conditions in the ISM. So far, many line survey observations toward characteristic sources in our Galaxy have been carried out using radio telescopes. For example, line surveys toward massive star forming regions were reported in Sgr B2 (Cummins et al. 1986; Friedel et al. 2004) and Ori-KL (Johansson et al. 1984; Turner 1989; 1991; Crockett et al. 2010). The results of the low-mass starless cores TMC-1 (cyanopolyyne peak) (Kaifu et al. 2004; Kalenskii et al. 2004) and the evolved star IRC+10216 (Avery et al. 1992; Kawaguchi et al. 1995; Cernicharo et al. 2010) were also published. Ultra compact HII regions in W51 and W3 (Bell et al. 1993; Helmich \& van Dishoeck 1997), and the shocked molecular cloud L1157 B1 (Sugimura et al. 2011; Yamaguchi et al. 2012) were observed. Moreover, Takekawa et al. (2014) reported the molecular composition toward the Galactic circumnuclear disk (CND) region and Sgr A*. The previous work showed that line surveys are useful not only in obtaining molecular abundances of known tracers, but also in discovering new chemical probes.

Over the last 10 years or so, targets for line surveys have been expanded to not only nearby Galactic sources but, with new instruments, sources with weaker emission from some external galaxies. These new instruments include low noise receivers with superconducting devices, wide band intermediate frequency chains, and wide band and high resolution spectrometers for large single-dish telescopes such as the Nobeyama Radio Observatory (NRO) 45-m telescope\footnote{The 45-m radio telescope is operated by Nobeyama Radio Observatory, a branch of the National Astronomical Observatory of Japan.} (e.g. Nakajima et al. 2008) and the Institut de Radioastronomie Millim\'{e}trique (IRAM) 30-m telescope (Carter et al. 2012). A pioneering spectral line survey toward the nearby galaxy NGC 253 was carried out by Mart\'{i}n et al. (2006) with the IRAM 30-m telescope in the 2-mm band. To date, a number of unbiased line surveys at mm/sub-mm wavelengths have been reported towards external galaxies using single-dish telescopes (e.g. Naylor et al. 2010; van der Werf et al. 2010; Aladro et al. 2011b, 2013, 2015; Snell et al. 2011; Costagliola et al. 2011; Kamenetzky et al. 2011; Spinoglio et al. 2012; Davis et al. 2013; Watanabe et al. 2014). Previous extragalactic line surveys were reviewed by Mart\'{i}n (2011). As a result, about 60 molecular species have been identified in nearby external galaxies (a list is available in CDMS\footnote{The Cologne Database for Molecular Spectroscopy (CDMS) can be accessed at http://www.astro.uni-koeln.de/cdms/.}; M\"{u}ller et al. 2005), and it is now possible to study molecular abundances and chemical reactions in external galaxies. In fact, some groups have suggested that it is possible to diagnose power sources in galactic nuclei using intensity ratios of molecular lines, such as HCN/HCO$^{+}$, HCN/CO, or HCN/CS (e.g. Jackson et al. 1993; Tacconi et al. 1994; Helfer \& Blitz 1995; Kohno et al. 1996, 2008; Usero et al. 2004; Imanishi et al. 2007; 2016; Krips et al. 2007; 2008; Davies et al. 2012; Krips 2012; Izumi et al. 2013; 2016). However recent high spatial resolution images of several molecules with the Atacama Large Millimeter/sub-millimeter Array (ALMA) toward NGC 1068 and NGC 1097 suggest that the interpretation of molecular lines at the vicinity of an active nucleus is still complicated (Garc\'{i}a-Burillo et al. 2014; Mart\'{i}n et al. 2015). In any case, it is important to investigate the relationships between the power sources of galaxies such as active galactic nuclei (AGNs) and/or starbursts, and the chemical properties of the surrounding dense interstellar medium. In order to study the effects on molecular abundances and to probe the power sources, it is needed to observe molecular lines in the irradiation by UV photons from starbursts or by X-rays from AGNs, which produce either a photon-dominated region (PDR) (e.g. Hollenbach \& Tielens 1999) or an X-ray dominated region (XDR) (e.g. Maloney et al. 1996; Lepp \& Dalgarno 1996; Meijerink \& Spaans 2005). It is expected to find different molecular signatures for these two regions.

Observations of the molecular gas chemistry of the AGN toward NGC 1068, one of the nearest galaxies with an AGN, have been reported by some groups (Usero et al. 2004; P\'{e}rez-Beaupuits et al. 2009; Garc\'{i}a-Burillo et al. 2010; Nakajima et al. 2011, 2015; Aladro et al. 2013, 2015; Takano et al. 2014; and Viti et al. 2014). However, the observed molecular lines in these previous works except for Aladro et al. (2013; 2015) and Takano et al. (2014) are limited to molecules of strong emission lines, such as CO, HCN, HCO$^{+}$, SiO, and CN. Therefore, further systematic observations of molecular lines using a compact beam without missing flux for seeing the affected molecular gas by the AGN are indispensable to study the impact of the AGN on the interstellar medium. Up to now, there have been two complete line survey observations, which have sufficient velocity resolution ($<$20 km s$^{-1}$) and high sensitivity ($\sim$1 mK) in the 3-mm band toward NGC 1068, one with the IRAM 30-m telescope by Aladro et al. (2013; 2015), and the other with the NRO 45-m telescope by Takano et al. (in prep., hereafter the data paper). The prior observations toward NGC 1068 by Snell et al. (2011) and Kamenetzky et al. (2011) are also line surveys at 74--111 GHz and 190--307 GHz, respectively. But the velocity resolution of their instruments, the Redshift Search Receiver (RSR) and a broadband millimeter-wave grating spectrometer (Z-Spec), is more than 100 km s$^{-1}$. Therefore, only strong lines with wide line widths were detected.

From the initial results of the line survey with the NRO 45-m telescope, we successfully detected basic carbon-containing molecules such as C$_{2}$H and cyclic-C$_{3}$H$_{2}$ (Nakajima et al. 2011). The relative abundances of these molecules with respect to CS were found to show no significant differences between NGC 1068 and NGC 253. Thus, it was concluded that these basic carbon-containing molecules are insensitive to the AGN conditions and/or that these molecules exist in a cold gas away from the AGN. Aladro et al. (2013) compared their results in NGC 1068 with the typical starburst galaxies NGC 253 and M 82, and they suggested that the gas in the nucleus of NGC 1068 has a different chemical composition from those of the starburst galaxies. In particular, they determined that the abundances of CN, SiO, HCO$^{+}$, and HCN are higher in NGC 1068. In contrast, H$_{2}$CO and CH$_{3}$CCH, which are abundant species in the starburst galaxies, are not detected in NGC 1068. The feature of abundant CH$_{3}$CCH in starburst galaxies was also suggested in Aladro et al. (2015), who obtained and compared the line survey spectra in the 3-mm band toward the starburst galaxies NGC 253, M 82, and M 83, the AGN-hosting galaxies M 51, NGC 1068, and NGC 7469, and the ultra-luminous infrared galaxies (ULIRGs) Arp 220 and Mrk 231.

In this paper, we report analysis results of molecular lines toward NGC 1068 based on the line survey in the 3-mm band using the NRO 45-m telescope. The beam size of this telescope ($\sim$19$^{\prime\prime}$ at 85 GHz) is smaller than the size of the starburst ring in NGC 1068 ($d$$\sim$30$^{\prime\prime}$). This small beam size is essential to study the impact of the AGN on the surrounding molecules, because it will enable us to separate the contamination of the molecular lines from the starburst region in NGC 1068. We also observed the typical starburst galaxies NGC 253 and IC 342 to compare the effects of the AGN and the starburst on molecular abundances. The details of observations and molecular spectra of each galaxy are presented in the data paper. In this paper, we present the results of the fractional abundances of all detected molecules, and discuss the physical and chemical properties, and possible scenarios for the differentiation of the molecular abundances in both types of galaxies. The analysis for the calculation of rotation temperatures and column densities is described in section 2, and these calculated results are presented in section 3. The trends found in comparing the molecular abundances in the three galactic nuclei and the possible explanations for particular molecular abundances are discussed in section 4.

\section{Analysis}

\subsection{Rotation Diagram}

We constructed rotation diagrams of the observed molecules in each galaxy. A list of detected and non-detected molecules is shown in Table 1. For molecules with detections of only one transition in our observations, we refer to the data of other transitions in other bands in previous papers. Although there are a lot of previous observations in the 3-mm band, we used here only our results in that frequency band in the rotation diagrams. Rotational temperatures ($T_{\rm rot}$) were calculated from the slopes of the fitted lines on the rotation diagrams under the assumptions that all lines are optically thin, that a single excitation temperature ($T_{\rm ex}$) characterizes all transitions, and that local thermodynamic equilibrium (LTE) pertains. We used the following equation (e.g., Turner 1991), 
\begin{equation}
{\rm log}\left(\frac{N_{\rm mol}}{Z}\right) = {\rm log}\frac{8{\pi}k{\nu}^{2}}{hc^{3}A_{ul}g_{u}g_{I}g_{K}}W + \frac{E_{u}}{k} \frac{{\rm log} \,e}{T_{\rm rot}},
\end{equation}
where $N_{\rm mol}$ is the column density, $Z$ the partition function, $\nu$ the frequency, $g_{u}$ the rotational degeneracy of the upper state (2$J_{u}$ + 1), $g_{I}$ and $g_{K}$ the reduced nuclear spin degeneracy and $K$-level degeneracy, respectively, and $E_{u}$ the energy of the upper state of the transition. 

The integrated intensity $W = \int T_{\rm mb} dv$ is corrected for the beam filling factor ($\eta_{\rm bf}$), which is calculated as $\eta_{\rm bf}$ = $\theta_{\rm s}^{2}$ / ($\theta_{\rm s}^{2}$ + $\theta_{\rm b}^{2}$). It accounts for the dilution effect due to the coupling between the source and the telescope beam in the approximation of a Gaussian source distribution of size $\theta_{\rm s}$ (FWHM) that is observed with a Gaussian beam of size $\theta_{\rm b}$ (HPBW). $\theta_{\rm b}$ of the receiver used on the NRO 45-m telescope was measured by cross scans of a point source emission such as a quasar (Nakajima et al. 2008). The adopted $\theta_{\rm b}$ for calculation of the beam filling factor is presented in the table in Appendix 2. The error of $\theta_{\rm b}$ obtained with the NRO 45-m telescope is about 2 \%, and the effect on the estimations of $T_{\rm rot}$ and $N_{\rm mol}$ can be negligible. The adopted $\theta_{\rm s}$ values of NGC 1068, NGC 253 and IC 342 are 4$^{\prime\prime}$, 20$^{\prime\prime}$ and 6$^{\prime\prime}$, respectively. They are taken from the distributions of HCN for NGC 1068 (Helfer \& Britz 1995), CS for NGC 253 (Peng et al. 1996) and HCN for IC 342 (Schinnerer et al. 2008), which are estimated from interferometric maps. Moreover, these $\theta_{\rm s}$ are similar to the values in the previous works for easy comparison (Bayet et al. 2009; Aladro et al. 2013; 2015 for NGC 1068, Mart\'{i}n et al. 2006 for NGC 253, and Bayet et al. 2009 for IC 342). Uncertainties of the error of $\pm$50 \% in the assumed source sizes translate into about $\pm$20 \% uncertainty in the estimates of $T_{\rm rot}$ and $N_{\rm mol}$.

$A_{ul}$ is the Einstein $A$-coefficient given by
\begin{equation}
A_{ul} = \frac{64{\pi}^{4}{\nu}^{3}}{3hc^{3}}\frac{S{\mu_{ul}}^{2}}{g_{u}},
\end{equation}
where $\mu_{ul}$ is transition dipole, and $S$ is the line strength. We obtain $\nu$ and $\mu_{ul}$ from the CDMS database. All rotation diagrams of the detected molecules in each galaxy are shown in Appendix 1, and the line parameters of the molecules are listed in Appendix 2. The upper or lower limits of the physical parameters for non-detected lines are calculated with their rms noise level (1 $\sigma$). It is noted that some molecular lines are optically thick, and the assumption of LTE is not valid. For example, the column densities of $^{12}$CO, HCN, and CN are possibly underestimated. The optical depths of these molecules are described in section 3.1.

\subsection{Column Density}

$N_{\rm mol}$ is calculated from the intercept at $E_{u}/k = 0$ of the fitted line on the rotation diagram and equation (1). The partition function $Z$ for each molecular species is calculated based on the following equation.
\begin{equation}
Z = \sum_{\rm all {\it E_{i}}}g_{u}g_{I}g_{K} {\rm exp}\left(\frac{-E_{i}}{kT_{\rm rot}}\right).
\end{equation}
We describe $Z$ for linear species, symmetric tops, and asymmetric tops in the next subsections based on Turner (1991). For polyatomic molecules except for linear species, we use the integral approximation for $Z$ because of the density of rotational energy levels. For the calculations of $N_{\rm mol}$ for molecules with only one detected transition, we assumed $T_{\rm rot}$ = 10$\pm$5 K, which is almost the same value of the obtained rotational temperatures of CS among the three galaxies in this observation (Table 2). Because the critical density of CS ({\it J}=2--1) is about 10$^{4}$--10$^{5}$ cm$^{-3}$, it traces the region where the densities are higher than that traced by CO, and this assumption of $T_{\rm rot}$ is probably reasonable for such molecules.

\subsubsection{Linear Species}
$Z$ for linear species is calculated numerically using the following equation, which we apply to $^{12}$CO, $^{13}$CO, C$^{18}$O, C$^{17}$O, CS, C$^{34}$S, HC$_{3}$N, SiO, HCN, H$^{13}$CN, HCO$^{+}$, H$^{13}$CO$^{+}$, HNC, N$_{2}$H$^{+}$, SO, CN, $^{13}$CN, and C$_{2}$H:
\begin{equation}
Z = \sum_{\rm {\it E_{i}}}g_{u} {\rm exp}\left(\frac{-E_{i}}{kT_{\rm rot}}\right).
\end{equation}
Note that $g_{I}$ and $g_{K}$ are equal to one for all levels in the case of the above molecules. 

\subsubsection{Symmetric Tops}
$Z$ for prolate symmetric tops is calculated using the following equation:
\begin{equation}
Z = \frac{1}{3}\left[\frac{\pi(kT_{\rm rot})^{3}}{h^{3} AB^{2}}\right]^{\frac{1}{2}}
\end{equation}
for molecules with C$_{3v}$ symmetry such as CH$_{3}$CN and CH$_{3}$CCH, while $A$ and $B$ are rotational constants. 

The line strength of CH$_{3}$CN and CH$_{3}$CCH are obtained from Boucher et al. (1980) and Bauer et al. (1979), respectively. For these molecules, $g_{K} = 1$ for $K = 0$ and $g_{K} = 2$ for $K \neq 0$, and $g_{I} = 1/2$ for $K = 3n$ and $g_{I} = 1/4$ for $K \neq 3n$ ($n = 0, 1, 2, \cdots$) are used in equation (1). Although the transitions of different $K$ladders are blended in our observations, we apply the method of separating for blended lines based on Mart\'{i}n et al. (2006).

\subsubsection{Asymmetric Tops}
$Z$ for asymmetric tops is calculated using the following equations:
\begin{equation}
Z = \left[\frac{\pi(kT_{\rm rot})^{3}}{h^{3} ABC}\right]^{\frac{1}{2}}
\end{equation}
for molecules with no symmetry such as HNCO, and
\begin{equation}
Z = \frac{1}{2}\left[\frac{\pi(kT_{\rm rot})^{3}}{h^{3} ABC}\right]^{\frac{1}{2}}
\end{equation}
for molecules with C$_{2v}$ symmetry such as cyclic-C$_{3}$H$_{2}$. For an internal rotor such as CH$_{3}$OH, we used the following equation for $A$-type and $E$-type species:
\begin{equation}
 Z(A) = Z(E) = \left[\frac{\pi(kT_{\rm rot})^{3}}{h^{3} ABC}\right]^{\frac{1}{2}},
\end{equation}
where $A$, $B$ and $C$ are rotational constants. 

The line strength of cyclic-C$_{3}$H$_{2}$ is obtained from Thaddeus et al. (1985) and Vrtilek et al. (1987), and that of CH$_{3}$OH is obatined from Lees et al. (1973). The $g_{I}$ is 3/4 for the ortho levels and 1/4 for the para levels for cyclic-C$_{3}$H$_{2}$. For CH$_{3}$OH, $g_{K} = 1$ and $g_{I} = 2$ for $A$-type, and $g_{K} = 2$ and $g_{I} = 1$ for $E$-type species are used.

\section{Results}
\subsection{Optical depth}

We calculated the optical depths of C$_{2}$H ({\it N} = 1--0), CN ({\it N} = 1--0), $^{13}$CN ({\it J} = 1--0), CS ({\it J} = 2--1), C$^{34}$S ({\it J} = 2--1), HCO$^{+}$ ({\it J} = 1--0), H$^{13}$CO$^{+}$ ({\it J} = 1--0), HCN ({\it J} = 1--0), H$^{13}$CN ({\it J} = 1--0), and the CO isotopic species ({\it J} = 1--0), and listed these values in Table 2. C$_{2}$H, CN, and $^{13}$CN show multiple line structures, which are caused by fine structure components. The intensities of each fine and hyperfine component were calculated based on spectroscopic theory and previously published experiments in the laboratory as discussed below. We can then obtain the optical depths from the intensity ratios between each component, and the total value of all components in the three galaxies are shown in Table 2. For C$_{2}$H, six Gaussian functions based on the expected intensity ratios with theory, as described by Tucker et al. (1974), were fitted to the C$_{2}$H spectra, as shown in figure 3. For CN (Skatrud et al. 1983) and $^{13}$CN (Bogey et al. 1984), nine and fourteen Gaussian functions, respectively, were fitted to each spectrum as shown in figures 4 and 5. The excitation temperature employed is the obtained rotational temperature or the assumed 10 K. The assumed cosmic background temperature is 2.7 K. For $^{12}$CO, $^{13}$CO, HCO$^{+}$, H$^{13}$CO$^{+}$, HCN and H$^{13}$CN, a carbon isotopic ratio [$^{12}$C]/[$^{13}$C] of 40 was assumed (Henkel et al. 2014). For C$^{18}$O and C$^{17}$O, the oxygen isotopic ratios [$^{16}$O]/[$^{18}$O] and [$^{16}$O]/[$^{17}$O] were assumed to be 145 and 1290, respectively (Henkel et al. 2014). Finally for CS and C$^{34}$S, the sulfur isotopic ratio [$^{32}$S]/[$^{34}$S] was assumed to be 23 (Wilson \& Matteucci 1992). Although previous observations of the sulfur isotopic ratio in external galaxies are limited, Frerking et al. (1980) suggested that the [$^{32}$S]/[$^{34}$S] ratio shows little or no variation at least in our Galaxy, such as in the local ISM, the Galactic center, and the solar system. Therefore, we applied the terrestrial sulfur isotopic ratio in this work.

The calculated optical depths of the observed lines of C$_{2}$H, $^{13}$CN, $^{13}$CO, C$^{18}$O, C$^{17}$O, C$^{34}$S, H$^{13}$CO$^{+}$, and H$^{13}$CN are smaller than unity, and we think that these are the optically thin lines. Care should be taken for CS, because its value is almost unity. Moreover, the observed lines of $^{12}$CO, HCN, and HCO$^{+}$ are optically thick in the beam size of the 45-m telescope. Therefore, the observed line intensities of these optically thick molecules are saturated depending on the optical depth, and the abundances based on the LTE assumption are underestimated. The optical depth of CN depends strongly on the galaxy. Those in NGC 1068 and IC 342 are more than unity but that in NGC 253 is 0.39. This value is more or less consistent with the value of $\sim$0.5 in NGC 253 by Henkel et al. (2014).

\subsection{Rotation Temperature and Column Density}

We calculated $T_{\rm rot}$ and $N_{\rm mol}$ for 23 molecular species in each galaxy. The derived values are listed in Table 3. The errors in $T_{\rm rot}$ and $N_{\rm mol}$ are calculated from the maximum and minimum of the slope and the intercept at the zero level of the energy of the fitting line, respectively, in the rotation diagram due to the errors of the integrated intensity. The errors of the integrated intensity for each line are described in the data paper. Note that for molecules with only one detected transition, such as cyclic-C$_{3}$H$_{2}$, C$_{2}$H, N$_{2}$H$^{+}$, CH$_{3}$OH, $^{13}$CN, and CH$_{3}$CN in NGC 1068, HNC, N$_{2}$H$^{+}$, CH$_{3}$OH, and $^{13}$CN in NGC 253, and cyclic-C$_{3}$H$_{2}$, H$^{13}$CN, H$^{13}$CO$^{+}$, SiO, C$_{2}$H, HNC, N$_{2}$H$^{+}$, and CH$_{3}$OH in IC 342, $T_{\rm rot}$ = 10$\pm$5 K is assumed. For H$^{13}$CN, C$^{34}$S, and SO in NGC 1068, and SO in IC 342, the upper or lower limit of $T_{\rm rot}$ and $N_{\rm mol}$ are calculated. Although these molecules were observed in two transitions, one transition was not detected and only the upper limit of the line intensity could be obtained. In addition, CH$_{3}$CCH in NGC 1068 and $^{13}$CN in IC 342 were not detected. Since C$^{17}$O emission in NGC 1068 is probably too weak to obtain a reasonable signal-to-noise ratio, C$^{17}$O is only marginally detected. Moreover, H$^{13}$CO$^{+}$ is not clearly detected in NGC 1068, because it is difficult to separate the weak H$^{13}$CO$^{+}$ line from the partially blended SiO spectrum.

We compare $N_{\rm mol}$ for each molecule among the three galaxies, and the results are shown in figure 1. As a result, although the molecular gas surrounding the CND is irradiated by high energy emission (e.g. X-rays and UV) from the AGN, the CND in NGC 1068 is found to be chemically very rich. The column densities of CN, HNCO, CS, HCN, HCO$^{+}$, HNC, HC$_{3}$N, $^{13}$CN, H$^{13}$CN, and SiO are clearly high, and on the contrary CH$_{3}$CCH is clearly low in column density toward NGC 1068 as compared with the starburst galaxies. The detailed features of each molecule are described below.

\subsubsection{$^{12}$CO and isotopic species}
There are many previous observations on $^{12}$CO, $^{13}$CO, and C$^{18}$O for all three galaxies, and we can plot more than three transitions for each CO isotopomer in the rotaion diagrams. The references of the previous observations are listed in the footnote of the tables in Appendix 2. The plots of $N_{u}/g_{u}$ of $^{12}$CO, $^{13}$CO, and C$^{18}$O ({\it J} = 1--0) with the IRAM 30-m in NGC 1068, which were reported by Aladro et al. (2013; 2015), are significantly high in comparison with the plots of our results with the NRO 45-m under the assumption of the same source size. As a result, the calculated $N_{\rm mol}$ of $^{12}$CO, $^{13}$CO and C$^{18}$O with the IRAM 30-m ((3.8$\pm$0.3)$\times$10$^{18}$ cm$^{-2}$, (4.6$\pm$0.1)$\times$10$^{17}$ cm$^{-2}$ and (1.29$\pm$0.04)$\times$10$^{17}$ cm$^{-2}$) are about twice as large as our results ((1.9$\pm$0.1)$\times$10$^{18}$ cm$^{-2}$, (1.2$\pm$0.0)$\times$10$^{17}$ cm$^{-2}$ and (5.3$^{+0.6}_{-0.5}$)$\times$10$^{16}$ cm$^{-2}$). These differences are likely to arise because the IRAM 30-m observations contain emission from the starburst ring in addition to the CND, given that the beam (HPBW$\sim$21$^{\prime\prime}$--29$^{\prime\prime}$) is larger than that of the 45-m telescope (HPBW$\sim$15$^{\prime\prime}$--19$^{\prime\prime}$). Molecular gas in NGC 1068 is distributed not only in the CND but also in the surrounding starburst ring region ($d\sim$20$^{\prime\prime}$--40$^{\prime\prime}$), which has been clearly seen with interferometric observations (e.g., Helfer \& Blitz 1995; Papadopoulos et al. 1996; Tacconi et al. 1997; Shinnerer et al. 2000; Garc\'{i}a-Burillo et al. 2014; Takano et al. 2014; Tosaki et al. 2017). This explanation is a possible reason for the finding of two components in the rotation diagram for at least $^{13}$CO by Aladro et al. (2013). Note that the $^{12}$CO opacity is evidently thick, and we tend to underestimate the values of $N_{u}$/$g_{u}$ in the rotation diagram especially at the lower energy level. The $T_{\rm rot}$ are determined to be approximately 20 K, 10 K, and 5 K for $^{12}$CO, $^{13}$CO, and C$^{18}$O, respectively, in NGC 1068 and IC 342. For NGC 253, $T_{\rm rot}$ for these molecules are about 10--20 K higher than in the other galaxies, but the $N_{\rm mol}$ are not higher. The C$^{17}$O lines were clearly detected in NGC 253 and IC 342 but marginally detected in NGC 1068. 

\subsubsection{CS and C$^{34}$S}
The CS {\it J} = 3--2, 5--4, and 7--6 transitions in NGC 1068 were reported by Mauersberger et al. (1989), Mart\'{i}n et al. (2009), and Bayet et al. (2009). However, Bayet et al. (2009) mentioned that the {\it J} = 7--6 line was only marginally detected. Subsequently, with the exception of the {\it J} = 7--6 line, the performed fit is shown in the rotation diagram in Appendix 1, and the obtained {\it T}$_{\rm rot}$ is 7.9$\pm$0.3 K. We fitted the data with only one spatial component in NGC 253 and IC 342, but it is probably better to fit with at least two components, as suggested by Bayet et al. (2009). In our observations toward NGC 1068, C$^{34}$S was not detected, and an upper limit of $N_{\rm mol}$ ($<$1.9$\times$10$^{13}$ cm$^{-2}$) was determined. On the other hand, the previous observations with the IRAM 30-m by Mart\'{i}n et al. (2009) and Aladro et al. (2013) detected C$^{34}$S {\it J} = 3--2 and {\it J} = 2--1, respectively. The derived column densities were (7.3$\pm$1.2)$\times$10$^{12}$ cm$^{-2}$ in Mart\'{i}n et al. (2009) and (1.3$\pm$0.2)$\times$10$^{14}$ cm$^{-2}$ in Aladro et al. (2015). This inconsistency with IRAM 30-m data may be caused in part by the fact that the assumed source sizes were not the same in these two studies. Mart\'{i}n et al. (2009) and Aladro et al. (2015) assumed source sizes of 10$^{\prime\prime}$ and 4$^{\prime\prime}$, respectively. If we compare our results with those of Aladro et al. (2015), which assumed the same source size as our study, the remaining source of the difference is that the emission of C$^{34}$S probably comes from not only the CND but also the starburst ring. In fact, we found that CS ({\it J} = 2--1) is distributed in the CND and also in the starburst ring with our ALMA observations (Takano et al. 2014).

\subsubsection{cyclic-C$_{3}$H$_{2}$ and C$_{2}$H}
 Because there are no previous observations of cyclic-C$_{3}$H$_{2}$ and C$_{2}$H except for the 3-mm band in NGC 1068 and IC 342, we assumed $T_{\rm rot}$ = 10$\pm$5 K for the calculations of the $N_{\rm mol}$. The $N_{\rm mol}$ values of cyclic-C$_{3}$H$_{2}$ and C$_{2}$H are (2.0$^{+1.7}_{-1.0}$)$\times$10$^{14}$ cm$^{-2}$ and (2.6$\pm$1.0)$\times$10$^{15}$ cm$^{-2}$, respectively in NGC 1068, and (6.9$^{+7.8}_{-4.3}$)$\times$10$^{13}$ cm$^{-2}$ and (5.7$^{+2.2}_{-2.1}$)$\times$10$^{14}$ cm$^{-2}$, respectively in IC 342. For NGC 253, the $N_{\rm mol}$ of these molecules were reported by Mart\'{i}n et al. (2006). Their obtained values are (3.0$\pm$6.0)$\times$10$^{13}$ cm$^{-2}$ and (1.2$\pm$0.1)$\times$10$^{15}$ cm$^{-2}$ for cyclic-C$_{3}$H$_{2}$ and C$_{2}$H, respectively, while our results are (4.4$\pm$0.0)$\times$10$^{13}$ cm$^{-2}$ and (1.7$\pm$0.0)$\times$10$^{15}$ cm$^{-2}$, and these values are consistent with their values. 

Moreover, we had already reported the $N_{\rm mol}$ of these molecules in NGC 1068 and NGC 253 in the paper containing the initial results of our line survey (Nakajima et al. 2011). These values based on the new observational data are not significantly changed from the initial work, as expected. But the $N_{\rm mol}$ of cyclic-C$_{3}$H$_{2}$ in NGC 1068 is changed from the initial result, because we did not take into account the beam filling factor for cyclic-C$_{3}$H$_{2}$ in this galaxy in Nakajima et al. (2011). Thus, the new $N_{\rm mol}$ of cyclic-C$_{3}$H$_{2}$ in NGC 1068 in this paper is likely to be more correct.

\subsubsection{HCN, HCO$^{+}$ and their $^{13}$C isotopic species}
HCN is one of the characteristic molecules of AGNs, because the intensity ratio of HCN with respect to HCO$^{+}$ or CO is often enhanced in AGN-dominant galaxies as compared with starburst-dominant galaxies (e.g. Kohno et al. 2008). However, HCN is abundant in galactic nuclei, and so easily becomes optically thick. Moreover, the spectra of HCN are complex because they have hyperfine structure and often show self-absorption effects. Therefore, calculations of temperature and column density of HCN accurately with the LTE assumption are difficult (e.g. Loughnane et al. 2012). In fact, the calculated optical depth of HCN from our observations is more than unity. We think H$^{13}$CN will be a useful candidate to investigate the HCN enhancemnent effect of AGNs. Unfortunately, so far the observations of H$^{13}$CN in NGC 1068 have been limited. We had already reported the ${\it J}$ = 1--0 line (Nakajima et al. 2011), and Wang et al. (2014) reported only an upper limit to the ${\it J}$ = 3--2 line with the Atacama Pathfinder Experiment (APEX). We obtained values for $N_{\rm mol}$ of HCN ((4.7$^{+0.0}_{-0.3}$)$\times$10$^{14}$ cm$^{-2}$) and H$^{13}$CN ($>$4.7$\times$10$^{13}$ cm$^{-2}$) in NGC 1068, which are approximately 2--4 times and $>$3 times higher than those in the starburst galaxies, respectively.

The HCO$^{+}$ (${\it J}$ = 1--0) line is often compared with the HCN (${\it J}$ = 1--0) line, because the rest frequencies of these lines are similar. Furthermore, the intensities of these lines are typically strong, and detections are facile. Because HCO$^{+}$ is also often optically thick, H$^{13}$CO$^{+}$ is also useful. But the frequency of the H$^{13}$CO$^{+}$ (${\it J}$ = 1--0) line is very close to that of SiO (${\it J}$ = 2--1), and therefore these lines are blended in NGC 1068, as shown in the observed spectra in the data paper. We could not separate these emission lines sufficiently to obtain line parameters. Therefore, only the upper limit ($<$4.5$\times$10$^{12}$ cm$^{-2}$) of $N_{\rm mol}$ of H$^{13}$CO$^{+}$ is estimated in this work.

\subsubsection{HNC}
HNC is abundant in dark molecular clouds and is deficient in high temperature regions (e.g. Hirota et al. 1998). On the other hand, it has been suggested that one of the causes of a bright HNC emission is the influence of UV radiation in PDRs and/or X-rays in XDRs at densities $n \lesssim$ 10$^{4}$ cm$^{-3}$ and at total column densities $N_{\rm H} >$ 3$\times$10$^{21}$ cm$^{-2}$ (Meijerink \& Spaans 2005). In our results, the $N_{\rm mol}$ of HNC in NGC 1068, (1.7$\pm$0.1)$\times$10$^{14}$ cm$^{-2}$, is approximately two times higher than those in the starburst galaxies, which are (8.7$^{+2.6}_{-1.7}$)$\times$10$^{13}$ cm$^{-2}$ and (7.6$^{+2.2}_{-1.4}$)$\times$10$^{13}$ cm$^{-2}$ in NGC 253 and IC 342, respectively. Probably, this relatively high column density is not due to low temperature but to the large amount of gas and/or effect of the strong emission in the CND. Aladro et al. (2015) reported that $N_{\rm mol}$ of HNC in NGC 1068 is about four times higher than in NGC 253, and this is consistent with our results.

\subsubsection{CN and $^{13}$CN}
CN is one of the key molecules used to study environments irradiated with extreme UV or X-rays, as mentioned in theoretical approaches (e.g. Lepp \& Dalgarno 1996; Meijerink \& Spaans 2005; Meijerink et al. 2007). According to these papers, an increased X-ray ionization rate in molecular clouds can enhance the abundance of CN with respect to that in PDRs. In the data paper, we found that the integrated intensity ratio of CN, which is calculated by the summation of all fine structure lines, divided by $^{13}$CO, is significantly higher in NGC 1068 ($\sim$5.2) than in NGC 253 ($\sim$2.5) and IC 342 ($\sim$1.3). The obtained $N_{\rm mol}$ of CN in NGC 1068 is 2--4 times higher than those of NGC 253 and IC 342 as might be expected. Note that the optical depth of CN is high as we discuss in section 3.1, and the $^{13}$C isotopic species of CN will be also important to study the CN abundance. In fact, it is remarkable that the $N_{\rm mol}$ of $^{13}$CN in NGC 1068 is an order of magnitude higher than those of the starburst galaxies. The values are (1.0$^{+0.3}_{-0.2}$)$\times$10$^{14}$ cm$^{-2}$, (1.2$^{+0.5}_{-0.3}$)$\times$10$^{13}$ cm$^{-2}$ and $<$4.4$\times$10$^{12}$ cm$^{-2}$ in NGC 1068, NGC 253 and IC 342, respectively. Although the line intensity is very weak and the observations are often difficult, it is possibly useful to trace material irradiated with strong UV and/or X-rays. Henkel et al. (2014) detected CN and $^{13}$CN in NGC 253 with the IRAM 30-m telescope. They reported that the CN excitation temperature and the column density are 3--11 K and 2$\times$10$^{15}$ cm$^{-2}$, respectively. These values in NGC 253 are consistent with our results in this paper.

\subsubsection{CH$_{3}$OH, SO, HNCO, \& CH$_{3}$CN}
CH$_{3}$OH, SO, and HNCO are typical shock-related species and are well-known probes in Galactic sources such as L1157, to study chemical and physical conditions (e.g. see Bachiller \& P\'{e}rez Guti\'{e}rrez 1997 for CH$_{3}$OH and SO, and Rodr\'{i}guez-Fern\'{a}ndez et al. 2010 for HNCO). CH$_{3}$CN is known to exist in hot cores (e.g., Churchwell et al. 1992; Crockett et al. 2015), and is regarded as a useful tracer to study the dense gas. Codella et al. (2009) confirmed the association of CH$_{3}$CN with gas affected by the passage of a shock wave. For CH$_{3}$CN in NGC 253, Mauersberger et al. (1991) and Mart\'{i}n et al. (2006) observed $J_{K}$ = 5$_{K}$--4$_{K}$, 7$_{K}$--6$_{K}$, 8$_{K}$--7$_{K}$, and 9$_{K}$--8$_{K}$, but we used only the highest transition line for the rotation diagram to better constrain the rotational temperature. We obtained the $N_{\rm mol}$ of CH$_{3}$OH in NGC 1068, NGC 253 and IC 342 to be (1.7$^{+0.3}_{-0.1}$)$\times$10$^{13}$ cm$^{-2}$, (1.7$^{+0.4}_{-0.1}$)$\times$10$^{13}$ cm$^{-2}$, and (1.9$^{+0.5}_{-0.1}$)$\times$10$^{13}$ cm$^{-2}$, respectively, and the $N_{\rm mol}$ of HNCO to be (6.8$^{+0.6}_{-0.1}$)$\times$10$^{14}$ cm$^{-2}$, (1.1$^{+0.1}_{-0.0}$)$\times$10$^{14}$ cm$^{-2}$, and (2.5$^{+0.1}_{-0.0}$)$\times$10$^{14}$ cm$^{-2}$ in each galaxy. As a result, the $N_{\rm mol}$ of HNCO in NGC 1068 is approximately 3--6 times higher than that in the starburst galaxies unlike CH$_{3}$OH. On the other hand, we have already observed these molecules in NGC 1068 with ALMA and reported their column densities (Takano et al. 2014; Nakajima et al. 2015). In Nakajima et al. (2015), the $N_{\rm mol}$ of CH$_{3}$OH and HNCO in the CND were measured to be 1.3 and 5.0 times higher than those in the starburst ring, respectively, under the assumption of the same $T_{\rm rot}$ in both of the regions. Therefore, we obtain a similar comparison between the abundances of CH$_{3}$OH and HNCO in the CND and the starburst environments using ALMA and the single dish telescope. Mart\'{i}n et al. (2009) and Aladro et al. (2013) reported the $T_{\rm rot}$ of HNCO in NGC 1068. Our result $T_{\rm rot}$ = 39.9$^{+9.5}_{-4.9}$ K is closer to that of Aladro et al. (2013), which is 29.7$\pm$0.1 K, rather than the result of 5.8$\pm$0.8 K by Mart\'{i}n et al. (2009).

\subsubsection{SiO}
It is well known that SiO is a good tracer of shocked gas (e.g. Mikami et al. 1992; Mart\'{i}n-Pintado et al. 1997), and we expect that this molecule is a useful tracer of shocks related to strong UV or X-ray irradiated regions. In fact, Garc\'{i}a-Burillo et al. (2010) suggested that the CND of NGC 1068 has a giant XDR based on the abundance ratios of SiO/CO and CN/CO seen with observation of high spatial resolution. Kelly et al. (2017) found a strong SiO peak to the east of the AGN and discussed shock events in the CND. In our observations, $N_{\rm mol}$ of SiO in NGC 1068 is 2.3--2.9 times higher than those of the starburst galaxies. Our result is almost consistent with that of Aladro et al. (2015), who reported the $N_{\rm mol}$ of SiO in NGC 1068 to be about 2.2 times higher than that in NGC 253.

\subsubsection{HC$_{3}$N}
We observed the HC$_{3}$N ${\it J}$ = 10--9, 11--10, and 12--11 lines in the 3-mm band. But the ${\it J}$ = 10--9 line toward NGC 1068 was not clearly detected, because this frequency range shows a baseline fluctuation as shown that spectrum in the data paper. Thus, we made the rotation diagram only with the ${\it J}$ = 11--10 and 12--11 lines. As a result, the $T_{\rm rot}$ and $N_{\rm mol}$ in NGC 1068 are 13.4$^{+32.2}_{-5.0}$ K and 1.7$^{+2.7}_{-0.9}\times$10$^{14}$ cm$^{-2}$, respectively. $N_{\rm mol}$ of HC$_{3}$N is approximately 5--6 times higher than those in NGC 253 and IC 342. But its error in NGC 1068 is appreciably large, and this overabundance of HC$_{3}$N must be taken with caution. The reason for this large error is the limited observed lines from only two closely-spaced energy levels.

\subsubsection{N$_{2}$H$^{+}$}
N$_{2}$H$^{+}$ is well known as a tracer of cold dense gas (e.g. Bergin et al. 2002). So far there are many observations of N$_{2}$H$^{+}$ in Galactic molecular clouds, but observations toward external galaxies are limited, especially in AGNs. In the previous observations in NGC 1068 with the NRAO 12-m and IRAM 30-m, the N$_{2}$H$^{+}$ (${\it J}$ = 1--0) was detected (Sage et al. 1995; Aladro et al. 2013). Aladro et al. (2015) reported that $N_{\rm mol}$ of N$_{2}$H$^{+}$ is (1.2$\pm$0.5)$\times$10$^{14}$ cm$^{-2}$, which is about three times higher than that in NGC 253. Our value is (5.8$^{+2.9}_{-1.8}$)$\times$10$^{13}$ cm$^{-2}$ in NGC 1068, and this is only 1.7--1.9 times higher than those of the starburst galaxies.

\subsubsection{CH$_{3}$CCH}
For CH$_{3}$CCH in NGC 253, Mauersberger et al. (1991) and Mart\'{i}n et al. (2006) observed $J_{K}$ = 8$_{K}$--7$_{K}$, 9$_{K}$--8$_{K}$, 10$_{K}$--9$_{K}$, and 13$_{K}$--12$_{K}$, but we used only the highest transition line for the rotation diagram to better constrain the rotational temperature as in the case with CH$_{3}$CN. CH$_{3}$CCH was not detected in NGC 1068 in our observations. We obtained an upper limit of $N_{\rm mol}$, $<$9.1$\times$10$^{13}$ cm$^{-2}$, and it is about a factor of 4 lower than in the starburst galaxies. The enhancement of CH$_{3}$CCH is a possible diagnostic tool of starbursts as opposed to AGNs. Aladro et al. (2015) reported that CH$_{3}$CCH emission lines are only seen in the starburst galaxies, and our observations support their results. The $T_{\rm rot}$ of CH$_{3}$CCH in NGC 253 and IC 342 are 40.5$\pm$0.2 K and 39.6$^{+0.9}_{-0.8}$ K, respectively, and those are almost the highest among all detected molecules. The reason for these high temperatures is that the dipole moment of CH$_{3}$CCH is relatively low (0.78 D; Burrell et al. 1980) sothat it is relatively easily thermalized (Aladro et al. 2011a). Inaddition, Guzm\'{a}n et al. (2014) obtained that the rotation temperatures of CH$_{3}$CCH in Galactic sources are about 55 K. They are consistent with our results and are the highest temperatures among their observed complex molecules.

\subsection{Column density ratios}

The integrated intensity ratios of HCN/HCO$^{+}$, HCN/CO or HCN/CS could be useful discriminators between AGN and starburst activity, as explained in Section 1. But we need to investigate whether the column density ratios of these molecules also show the same trend as the intensity ratios. Several important column density ratios in our observations are listed in Table 4. The ratios of $\frac{N\rm{(HCN)}}{N\rm{(HCO^{+})}}$ are 2.2$^{+0.1}_{-0.2}$, 2.0$^{+0.6}_{-0.8}$, and 1.6$\pm0.1$ in NGC 1068, NGC 253, and IC 342, respectively, while the values of $\frac{N\rm{(HCN)}}{N\rm{(CS)}}$ are 0.89$^{+0.07}_{-0.06}$, 0.87$^{+0.04}_{-0.08}$, and 0.57$^{+0.07}_{-0.09}$ in the same galaxies. We found no clear enhancement of the HCN column density relative to the HCO$^{+}$ or CS column densities in the galaxy hosting an AGN compared with the starburst galaxies on a 1 kpc scale. However, the corresponding ratios of $\frac{N\rm{(H^{13}CN)}}{N\rm{(H^{13}CO^{+})}}$ become $>$10.4, 4.8$^{+1.0}_{-0.6}$, and 3.6$\pm1.5$, while the $\frac{N\rm{(H^{13}CN)}}{N\rm{(CS)}}$ ratios are $>$0.09, 0.05$\pm0.01$, and 0.07$^{+0.01}_{-0.14}$. Although the difference of $\frac{N\rm{(H^{13}CN)}}{N\rm{(CS)}}$ between NGC 1068 and IC 342 is small, its difference from NGC 253 and $\frac{N\rm{(H^{13}CN)}}{N\rm{(H^{13}CO^{+})}}$ clearly shows the enhancement in the galaxy hosting AGN (figure 2). Observations of the lines of H$^{13}$CN and H$^{13}$CO$^{+}$, which are optically thin unlike the lines of HCN and HCO$^{+}$, in the nucleus of galaxies will be more useful for differentiation, though high sensitivity observations are necessary.

CN and $^{13}$CN, which are produced in PDRs but also particularly in XDRs as described in section 3.2.6, are expected to be enhanced in the galaxies hosting AGNs compared with starburst galaxies. In fact, the CN line intensities are thought to show enhancement in extremely X-ray irradiated environments (e.g. Aalto et al. 2002), and we confirmed these features as column density ratios. The ratios of $\frac{N\rm{(CN)}}{N\rm{(CS)}}$ are 5.7$^{+0.6}_{-0.4}$, 4.8$^{+0.0}_{-0.4}$, and 3.9$^{+0.4}_{-0.6}$ in NGC 1068, NGC 253, and IC 342, respectively. But the ratio of $\frac{N\rm{(CN)}}{N\rm{(CS)}}$ in the galaxy hosting an AGN is a factor of only 1.2--1.5 larger than those in the starburst galaxies. On the other hand, the $^{13}$C isotopic species show a clearer enhancement in the galaxy hosting an AGN. The ratios of $\frac{N\rm{(^{13}CN)}}{N\rm{(CS)}}$ are 0.19$^{+0.06}_{-0.04}$, 0.05$^{+0.02}_{-0.01}$, and $<$0.02 in NGC 1068, NGC 253, and IC 342, respectively. This ratio is 4--10 times larger in NGC 1068 than those in the starburst galaxies. These results are probably due to a difference in the optical depth because the $^{13}$C species reveal the difference in the physical conditions deep inside of the molecular clouds, which may not be probed with the optically thick normal species.

The CN/HCN column density ratio is expected to probe radiation (UV-photon / X-rays) of the ISM in theory as discussed in 3.2.6. Meijerink et al. (2007) suggested that the $\frac{N\rm{(CN)}}{N\rm{(HCN)}}$ ratio is expected to be larger than unity in PDRs, but particularly large in XDRs. In our results, $\frac{N\rm{(CN)}}{N\rm{(HCN)}}$ is 6.4$^{+0.4}_{-0.0}$, 5.5$^{+0.3}_{-0.0}$, and 6.8$\pm$0.6. Although we should not forget concerns about the CN and HCN opacities, these ratios are almost the same among these three galaxies, and do not allow us to distinguish between the AGN and starbursts. On the other hand, the ratio of the $^{13}$C isotopic species $\frac{N\rm{(^{13}CN)}}{N\rm{(H^{13}CN)}}$ are $<$2.1, 1.0$^{+0.5}_{-0.3}$ and $<$0.3. The upper limit in NGC 1068 is about twice the actual values in NGC 253 and IC 342.

Finally, we show the ratio of $\frac{N\rm{(CH_{3}CCH)}}{N\rm{(C^{18}O)}}$ in Table 4. This ratio has already been reported by Aladro et al. (2015) in the line survey with the IRAM 30-m telescope (see Table 4 in their paper). They reported that CH$_{3}$CCH was not detected toward the galaxies hosting an AGN among eight observed various galaxies, and the ratio in NGC 1068 is less than about half of that in NGC 253. In our observations, the upper limit of this ratio in NGC 1068 is 7 times lower than those ratios in the starburst galaxies, and these results are consistent with the observations using the IRAM 30-m. Moreover, the observations with the NRO 45-m telescope place a more stringent constraint on the low-abundance of CH$_{3}$CCH in NGC 1068.

\section{Discussion}

\subsection{Trend of Molecular Abundances surrounding the AGN}
Comparison among the molecular abundances in each galaxy are shown in figures 6 and 7. In these figures we calculated the fractional abundances with respect to $^{13}$CO or CS. Although the fundamental fractional abundance is relative to molecular hydrogen ($N_{\rm H_{2}}$), we take a ratio of the column density of each species over the $^{13}$CO or CS column density in these figures.  We do this because an estimate of $N_{\rm H_{2}}$ is quite difficult based on CO line measurment using an assumed so called $X_{\rm CO}$ factor, which is a CO-to-H$_{2}$ conversion factor (e.g. Dickman 1978). The $X_{\rm CO}$ factor is easily affected depending on the physical conditions in the interstellar molecular gas. On the one hand, the obtained column densities of $^{13}$CO in our obeservations are not significantly different between the AGN and the starburst galaxies, and in addition their errors are the smallest among all detected molecules in our observations. The difference of $N$($^{13}$CO) between the AGN and the starbursts is within a factor of 1.4, and this is the smallest variation among all detected molecules as shown in figure 1 and Table 3. Therefore, we would expect to clearly see the effect of an AGN to compare with starbursts using the fractional abundances relative to $^{13}$CO.  However, we found that the $^{13}$CO $J$ = 1--0 line in the CND toward NGC 1068 is extremely weaker than that in the starburst ring from the observations with higher spatial resolution using ALMA (Takano et al. 2014). Although this signature is not seen with the single-dish telescope observations, a number of determined fractional abundances relative to $^{13}$CO may show excessive enhancement in NGC 1068. Therefore, we also calculate the fractional abundances with respect to CS, which is known to show little variation in abundance among galaxies with different activity  such as starbursts, AGNs, ULIRGs, and normal galaxies (Mart\'{i}n et al. 2009). In fact, the difference of $N$(CS) between the AGN and the starbursts in our results is within a factor of 2.5, and this value is one of the smallest variations next to CO and its isotopic species. In addition, the critical density of CS ($J$ = 2--1) is about 10$^{4-5}$ cm$^{-3}$, and therefore it traces regions where the densities are higher than that traced by $^{13}$CO. It is much more reasonable for comparisons involving molecules of high critical density.

Figure 6 shows correlation plots of the fractional abundances relative to (a) $^{13}$CO and (b) CS between NGC 253 and IC 342, which are both starburst galaxies. All differences of the fractional abundance are within an order of magnitude, and a good correlation of the abundances can be seen between these two galaxies. The correlation coefficients between these starburst galaxies are 0.996 for both $^{13}$CO and CS. This result suggests that average environments causing these chemical compositions are similar within the beam size of the 45-m telescope. 

Figure 7 contains correlation plots for NGC 1068 against both NGC 253 and IC 342 for the fractional abundances of the assorted molecules normalized by $^{13}$CO and CS. These are roughly in good correlation between the galaxy with an AGN and the starburst galaxies. In addition, we found that some molecules show significantly higher or lower fractional abundances in NGC 1068 relative to both of the starburst galaxies. It is interesting to note that such features among these galaxies are common. Molecules of high fractional abundance in NGC 1068 include CN, $^{13}$CN, HCN, H$^{13}$CN, and HC$_{3}$N, while CH$_{3}$CCH is deficient in this galaxy.

Figure 8 shows more clearly the ratios of the fractional abundances of assorted species with respect to $^{13}$CO and CS between NGC 1068 and the starburst galaxies. In these figures, the molecules are arranged in descending order of the abundance ratio from the left side on the horizontal axis. Therefore, the fractional abundance ratios greater than unity are on the left side of the figures, where the more abundant molecules in NGC 1068 are located. The highest ratio occurs for $^{13}$CN, for which it is greater than or equal to 10, while the lowest ratio is for CH$_{3}$CCH. The basic carbon-containing molecules, cyclic-C$_{3}$H$_{2}$ and C$_{2}$H, show similar abundances with respect to CS within factors of a few among the three galaxies. Therefore, the conclusion of our initial results (Nakajima et al. 2011) that the abundances are insensitive to the effect of the AGN for these molecules and/or that these molecules exist in a cold gas away from the AGN, is supported.

So far, there have been many observations of HCN and CN in NGC 1068 with single-dish telescopes (e.g. Nguyen-Q-Rieu et al. 1992; Paglione et al. 1997; Perez-Beaupuits et al. 2007, 2009; Krips et al. 2008; Kamenetzky et al. 2011; Aladro et al. 2013, 2015; and Wang et al. 2014). The line intensities of HCN and CN are higher than most of the molecules in these previous observations. Aladro et al. (2015) reported the column densities of HCN, CN and CO isotopic species in NGC 1068 as well as those in four starburst galaxies: NGC 253, M 82, M 83, and Arp 220. We calculated the ratios of $\frac{N\rm{(HCN)}}{N\rm{(^{13}CO)}}$ and $\frac{N\rm{(CN)}}{N\rm{(^{13}CO)}}$ in NGC 1068 based on their reported column densities, and these values are 2.3 and 1.9 times, respectively higher than the averaged ratios among those four starburst galaxies. On the other hand, the ratios of $\frac{N\rm{(HCN)}}{N\rm{(^{13}CO)}}$ and $\frac{N\rm{(CN)}}{N\rm{(^{13}CO)}}$ in NGC 1068 in our results are 3.7 and 3.9 times higher than the averaged ratios in the two starburst galaxies NGC 253 and IC 342. These higher ratios than those in Aladro et al. (2015) imply that the observations with the NRO 45-m telescope, the beam size of which is smaller than the IRAM 30-m telescope, tend to obtain molecular abundances in the CND with small contributions of emission lines from the starburst ring region.

\subsection{Possible Scenario for the Differentiation of Molecular Abundances}

We found that CH$_{3}$CCH is deficient, and that CN, $^{13}$CN, HCN, H$^{13}$CN, and HC$_{3}$N are abundant in the galaxy hosting an AGN, NGC 1068, relative to the starburst galaxies on a 1 kpc scale with the 45-m telescope. We discuss a possible scenario for the difference in these molecular abundances.

\subsubsection{CH$_{3}$CCH and HC$_{3}$N}
The low abundance of CH$_{3}$CCH relative to H$_{2}$, C$^{34}$S, and C$^{18}$O in NGC 1068 compared with those in the starburst galaxies NGC 253, M 82 and M83 obtained with the IRAM 30-m observations were already reported by Aladro et al. (2011a; 2011b; 2013; 2015). They obtained that the $\frac{N\rm{(CH_{3}CCH)}}{N\rm{(C^{18}O)}}$ ratios are $\gtrsim$2, $\gtrsim$10 and $\gtrsim$2 times more abundant in NGC 253, M 82 and M 83, respectively, than that in NGC 1068. They claimed that the fractional abundances of CH$_{3}$CCH are likely to be intrinsically related to the different nuclear activities in galaxies, implying that the chemistry of this molecule is different between starbursts and AGNs. $\frac{N\rm{(CH_{3}CCH)}}{N\rm{(C^{18}O)}}$ ratios in NGC 253 and IC 342 based on our results are more than $\gtrsim$5 times higher than that in NGC 1068, as can be seen in Table 4. Such a tendency is also seen in the $\frac{N\rm{(CH_{3}CCH)}}{N\rm{(CS)}}$ ratios in our results, as shown in figures 7 and 8. Therefore, our observations using a more compact beam strongly support the previous observational studies.

As shown in figure 8, the fractional abundance of HC$_{3}$N in NGC 1068 is higher than that in the starburst galaxies. In particular, the abundance of HC$_{3}$N with respect to $^{13}$CO in NGC 1068 is 6--9 times higher than those in NGC 253 and IC 342. But the error bar of the column density of HC$_{3}$N in NGC 1068 is quite large due to the low quality of the observational data with the 45-m telescope, and we think it may not be an accurate feature, as previously discussed in section 3.2.9. In fact, such a feature of HC$_{3}$N in galaxies hosting an AGN is not seen in some previous studies (e.g. Costagliola et al. 2011; Aladro et al. 2013; 2015). However, we clearly detected HC$_{3}$N $J$ = 11--10 and 12--11 lines in the CND toward NGC 1068 with ALMA (Takano et al. 2014), and we obtained that the abundance of HC$_{3}$N is approximately 30 times higher than that in the starburst ring region in NGC 1068 (Nakajima et al. 2015). Therefore, the tendency towards a relatively high abundance of HC$_{3}$N in the CND in NGC 1068 is already confirmed with ALMA observations. 

According to model calculations, CH$_{3}$CCH and HC$_{3}$N are both easily dissociated by UV (PDRs) and/or X-ray radiations (XDRs) (Aladro et al. 2013). Model calculations using a gas-phase network including high-temperature reactions (Harada et al. 2010) show that fractional abundances of HC$_{3}$N and CH$_{3}$CCH both increase with temperature. At the same time, the fractional abundance of CH$_{3}$CCH produced in a high-temperature environment is only $\sim$10$^{-9}$, much lower than that of HC$_{3}$N ($\sim$10$^{-7}$). On the other hand, there are some production routes of CH$_{3}$CCH involving grain reactions. For example, a precursor of CH$_{3}$CCH, C$_{2}$H$_{4}$ can be made efficiently on grains. When C$_{2}$H$_{4}$ desorbs from grains, the following gas-phase reaction forms CH$_{3}$CCH: ${\rm CH + C_{2}H_{4} \longrightarrow CH_{3}CCH + H}$ (Miettinen et al. 2006). Our calculations with a gas-grain code (Hersant et al. 2009) also show that CH$_{3}$CCH on ice can be made in high abundance by hydrogenation of cyclic-C$_{3}$H$_{2}$. As noted previously, the regions shielded from strong radiation are needed to help these molecules survive. We claimed that the emission of HC$_{3}$N in the CND in NGC 1068 must be coming from regions shielded from X-rays (Harada et al. 2013; Nakajima et al. 2015), but the reason for the non-detection of CH$_{3}$CCH in NGC 1068 is not clearly understood on $\sim$kpc scale with the single-dish observations. In contrast, Costagliola et al. (2015) reported the possibility of both HC$_{3}$N and CH$_{3}$CCH enhancement toward NGC 4418, which is a galaxy with a very compact and luminous nucleus, as seen with ALMA. We think that the differences in the fractional abundances between these molecules among the galaxies may be caused by a difference in the type and/or frequency of shocks, or by another heating mechanism that affects the synthesis of these molecules in active galactic nuclei. In any case, observations of extragalactic CH$_{3}$CCH are limitted so far, and relationships between active nuclei and its abundances are still uncertain.

\subsubsection{CN and HCN}
The radical CN and its isotopic species $^{13}$CN are more abundant in NGC 1068 compared with the starburst galaxies, as shown in figures 7 and 8. In particular, the ratio of $\frac{N\rm{(CN)}}{N\rm{(CS)}}$ in NGC 1068 is approximately 1.2--1.5 times higher than those in the starburst galaxies. Furthermore, $\frac{N\rm{(^{13}CN)}}{N\rm{(CS)}}$ is from 3.8 times to an order of magnitude larger in NGC 1068 than in the starburst galaxies as already shown in figure 2. This CN enhancement is also found in the high-spatial resolution observations toward NGC 1068 with ALMA (Nakajima et al. 2015). We reported the comparison of $\frac{N\rm{(CN)}}{N\rm{(CS)}}$ between the central $\sim$100 pc of the CND and the starburst ring region in NGC 1068, and showed that ratio in the CND is approximately 7.2 times higher than in the starburst ring region. This suggests that the reason for CN enhancement in NGC 1068 is a high abundance of CN in the CND.

Both chemical models and observations suggest that the CN fractional abundance is high when the chemistry is at an early stage of evolution of molecular clouds (Le Petit et al. 2006) or when there is high flux of UV photons, X-rays, and/or cosmic rays to ionize and dissociate precursor molecules such as HCN (Jansen et al. 1995; Lepp \& Dalgarno 1996; Hirota et al. 1999; Meijerink et al. 2007; Harada et al. 2013). In addition, Harada et al. (2013) suggested that CN is not increased by mechanical heating such as turbulence/shock heating. In our observations, HCN and H$^{13}$CN in NGC 1068 are also abundant, as shown in figure 8. The ratios $\frac{N\rm{(HCN)}}{N\rm{(CS)}}$ and $\frac{N\rm{(H^{13}CN)}}{N\rm{(CS)}}$ in NGC 1068 are approximately 1.0--1.6 times and more than 1.8 times, respectively, larger than in the starburst galaxies. Note that the ratios of $\frac{N\rm{(CN)}}{N\rm{(CS)}}$ and $\frac{N\rm{(HCN)}}{N\rm{(CS)}}$ are not so high compared with the starburst galaxies as those of $\frac{N\rm{(^{13}CN)}}{N\rm{(CS)}}$ and $\frac{N\rm{(H^{13}CN)}}{N\rm{(CS)}}$, but this reason is likely the high optical depths of CN and HCN as already discussed in section 3.3. The optically thin lines of $^{13}$CN and H$^{13}$CN are more useful for this discussion.

Meijerink et al. (2007) suggested that the range of $\frac{N\rm{(CN)}}{N\rm{(HCN)}}$ lies between 40 (at $n\sim$10$^{6}$ cm$^{-3}$) to over 1000 (at $n\sim$10$^{4}$ cm$^{-3}$) in an XDR based on their calculations. On the other hand, we obtained $\frac{N\rm{(CN)}}{N\rm{(HCN)}}$ and $\frac{N\rm{(^{13}CN)}}{N\rm{(H^{13}CN)}}$ ratios of 6.4 and 2.1, respectively in NGC 1068, whereas Aladro et al. (2015) reported that these values are 2.8 and $<$1.9, respectively. Although small differences in these ratios among these observations are probably caused by the difference of the beam sizes, these ratios are significantly low relative to a pure XDR environment. In fact, P\'{e}rez-Beaupuits et al. (2009) suggested that $\frac{N\rm{(CN)}}{N\rm{(HCN)}}$ in NGC 1068, obtained from the observations with the James Clerk Maxwell Telescope (JCMT), is relatively low ($<$ 1) with respect to what would be expected in a pure XDR scenario. Therefore, these abundance ratios are not explained only by the XDR scenario, because $\frac{N\rm{(CN)}}{N\rm{(HCN)}}$ as well as $\frac{N\rm{(^{13}CN)}}{N\rm{(H^{13}CN)}}$ are small ratios which are unexpected in a pure XDR environment.

In hot cores, a high temperature reaction ${\rm CN + H_{2} \longrightarrow HCN + H}$ efficiently converts CN into HCN reducing the fractional abundance of CN at an elevated temperature and increasing that of HCN (Harada et al. 2010). Therefore, one possible scenario for the high abundance of both CN and HCN in the CND in NGC 1068 is that the abundances of CN, $^{13}$CN are increased by X-ray and/or cosmic ray irradiation, while HCN and H$^{13}$CN are increased by mechanical heating due to turbulance and shock such as an AGN jet. This scenario has already been reported in another AGN, NGC 1097, by Izumi et al. (2013). In fact, Garc\'{i}a-Burillo et al. (2014) identified the signature of an AGN-driven outflow in the CND of NGC 1068 based on the CO velocity field.

Beam sizes of single-dish observations in our study and previous studies correspond to $\sim$1 kpc. If X-rays are not attenuated, they can cause a significant ionization rate even at this scale (e.g., Eq. 4 in Maloney et al. 1996). However, X-rays can be attenuated by a column density $N_{\rm H}\sim$10$^{24}$ cm$^{-2}$, and it is likely that they are attenuated in a 10-pc or 100-pc scale within the CND. Generally single-dish telescopes observe the whole structure of molecular clouds, which are both irradiated by strong radiations and shielded from them. Therefore, it is difficult to compare the observations and the model calculations. If we discuss the relationship between such physical phenomena and chemical characteristics in NGC 1068, we need to clearly separate the molecular abundances in the CND and the starburst ring. 

Finally, metallicity or elemental abundance is one of the important factors for the molecular abundance. While the galaxies have a metallicity gradient in general, the metallicities in NGC 253 and IC 342 are known to be close to the solar value, $Z$$\sim$$Z_{\odot}$ (Webstar \& Smith 1983; Yamagishi et al. 2011), and nearly solar, 0.8 $Z_{\odot}$ (Crosthwaite et al. 2001), respectively. Although the metallicity of AGNs are generally high, for example the values of super solar ($\gtrsim$3 $Z_{\odot}$) especially in the narrow line region (NLR) (e.g. Groves et al. 2006), all elements are consistent with solar value except for Nitrogen (N/H) in NGC 1068 (Kramer et al. 2015, Martins et al. 2010, Brinkman et al. 2002, Marshall et al. 1993). As a result, we judged that the metallicities in our objects observed with our beam size of $\sim$19$^{\prime\prime}$ are almost consistent with the solar value. Therefore, the difference of metallicity is likely to be not a main cause of the enhancement of some species such as HCN (at least in our observations with large beam size). The significant effect of metallicity is likely to be possible closer to the AGN core.

In order to investigate the effect of the AGN on the surrounding molecular gas, it is important to obtain the distribution of molecules in the CND with much higher spatial resolution. However, the investigation of molecular composition in nearby galaxies with single-dish telescopes is important. For example, it is necessary to observe molecular gas in high-redshift galaxies with interferometry for understanding the scenario of galaxy evolution. But these galaxies are not well resolved in their inner structure even with ALMA. At that time, the results of line surveys with single-dish telescopes will be the most basic template.

\section{Conclusions}

We carried out unbiased molecular line survey observations in the 3-mm band toward NGC 1068, NGC 253, and IC 342 with the NRO 45-m telescope. The rotation temperatures and column densities for 23 molecular species in each galaxy were obtained and added to other transition data in previous studies under the assumption of local thermodynamic equilibrium. In order to investigate the effect of an AGN on the surrounding interstellar medium, we determined molecular fractional abundances relative to $^{13}$CO and CS, and compared these abundances between NGC 1068, the galaxy hosting an AGN, and two starburst galaxies with correlation plots. The main results of this work are summarized as follows:\\

\begin{enumerate}
\item We found that the fractional abundances of the assorted molecules normalized by $^{13}$CO and CS are roughly in good correlation between NGC 1068 and the starburst galaxies NGC 253 and IC 342. This means that significant effects on the chemistry due to the existence of an AGN or starburst are not prominently seen over a large scale ($\sim$1 kpc) with the NRO 45-m telescope. But some molecules show higher or lower abundances in NGC 1068 relative to the starburst galaxies.

\item The molecules with higher fractional abundances in NGC 1068 include HCN, H$^{13}$CN, CN, $^{13}$CN and HC$_{3}$N, while only CH$_{3}$CCH shows a serious deficiency in its fractional abundance in NGC 1068. Although these trends are almost consistent with a prior line survey using the IRAM 30-m telescope, our results with the NRO 45-m telescope place more stringent constraints on the trend of the molecular abundances in NGC 1068.

\item  The radical CN and its isotopic species $^{13}$CN are clearly more abundant in NGC1068 compared with the starburst galaxies. In addition, HCN and H$^{13}$CN in NGC 1068 are also more abundant in our observations. We successfully detected the $^{13}$C isotopic species of CN and HCN with more than 3-sigma, and obtained their abundances in NGC 1068. The optically thin lines of $^{13}$CN and H$^{13}$CN are more useful for studies of the CN and HCN abundances.

\item  The obtained $\frac{N\rm{(CN)}}{N\rm{(HCN)}}$ and $\frac{N\rm{(^{13}CN)}}{N\rm{(H^{13}CN)}}$ ratios are 6.4 and 2.1, respectively in NGC 1068, and these values are relatively low with respect to what would be expected in chemical models of a pure XDR environment. One possible scenario for the high abundances of both CN and HCN in the CND is that the fractional abundances of CN and $^{13}$CN are increased by X-ray and/or cosmic ray irradiation. On the other hand, HCN and H$^{13}$CN are possibly increased by mechanical heating due to turbulence and shocks such as an AGN jet.

\item The reason for the non-detection of CH$_{3}$CCH in NGC 1068 is likely to be dissociation by high energy radiation, a lack of formation on grains, or less sublimation of a precursor of CH$_{3}$CCH, C$_{2}$H$_{4}$, from grains. On the contrary, the fractional abundance of HC$_{3}$N in NGC 1068 is found to be higher than those in the starburst galaxies. The emission of HC$_{3}$N must be coming from regions shielded from X-rays in the CND, and these regions are expected to be in a high temperature environment. This senario is consistent with the results from abundances of HCN and CN.

\end{enumerate}

\bigskip

This work is based on observations with the NRO 45-m telescope, as a part of the line surveys of the legacy projects. The authors thank the project members for helpful discussion. E. H. thanks the National Science Foundation (US) for their support of his program in astrochemistry. We also thank all of the staff of the 45-m telescope for their support. We are also grateful to Ryohei Kawabe for his advice and support of the line survey project. 

\clearpage

\begin{table}
  \caption{Detected and non-detected molecules in the 3-mm band.}\label{}
  \begin{center}
    \begin{tabular}{cccc}
      \hline
      Molecule & NGC 1068 & NGC 253 & IC 342 \\
      \hline
Linear species &&&\\
$^{12}$CO & Yes & Yes & Yes \\
$^{13}$CO & Yes & Yes & Yes \\
C$^{18}$O & Yes & Yes & Yes \\
C$^{17}$O & No & Yes & Yes \\
CS & Yes & Yes & Yes \\
C$^{34}$S & No & Yes & Yes \\
HC$_{3}$N & Yes & Yes & Yes \\
SiO & Yes & Yes & Yes \\
HCN & Yes & Yes & Yes \\
H$^{13}$CN & Yes & Yes & Yes \\
HCO$^{+}$ & Yes & Yes & Yes \\
H$^{13}$CO$^{+}$ & No & Yes & Yes \\
HNC & Yes & Yes & Yes \\
N$_{2}$H$^{+}$ & Yes & Yes & Yes \\
SO & Yes & Yes & Yes \\
CN & Yes & Yes & Yes \\
$^{13}$CN & Yes & Yes & No \\
C$_{2}$H & Yes & Yes & Yes \\
Symmetric tops &&&\\ 
CH$_{3}$CN & Yes & Yes & Yes \\
CH$_{3}$CCH & No & Yes & Yes \\
Asymmetric tops &&&\\
HNCO & Yes & Yes & Yes \\
CH$_{3}$OH & Yes & Yes & Yes \\
cyclic-C$_{3}$H$_{2}$ & Yes & Yes & Yes \\
      \hline
      \multicolumn{4}{@{}l@{}}{\hbox to 0pt{\parbox{75mm}{\footnotesize
Yes and No represent the detected and non-detected molecule, respectively, in our line survey observation with the NRO 45-m telescope. The observed parameter of molecular lines in each galaxy will be reported in the data paper (Takano et al. in prep.).
     }\hss}}
    \end{tabular}
  \end{center}
\end{table}

\begin{table}
 \tbl{Optical depth.}{%
 \begin{tabular}{lccc}
  \hline
  Molecule & NGC 1068 & NGC 253 & IC 342 \\
  \hline
  C$_{2}$H & 0.09 & 0.10 & 0.10 \\
  CN & 1.42 & 0.39 & 4.09 \\
  $^{13}$CN & 0.01 & 0.01 & ---\footnotemark[$*$] \\
  $^{12}$CO & 2.06 & 2.68 & 4.20 \\
  $^{13}$CO & 0.05 & 0.07 & 0.11 \\
  C$^{18}$O & 0.02 & 0.02 & 0.03 \\
  C$^{17}$O & ---\footnotemark[$*$] & 0.001 & 0.004 \\
  CS & ---\footnotemark[$*$] & 0.91 & 1.28 \\
  C$^{34}$S & ---\footnotemark[$*$] & 0.04 & 0.06 \\
  HCO$^{+}$ & ---\footnotemark[$*$] & 3.81 & 1.24 \\
  H$^{13}$CO$^{+}$ & ---\footnotemark[$*$] & 0.08 & 0.02 \\
  HCN & 2.49 & 3.09 & 3.46 \\
  H$^{13}$CN & 0.05 & 0.06 & 0.07 \\
  \hline
  \multicolumn{4}{@{}l@{}}{\hbox to 0pt{\parbox{60mm}{\footnotesize
    \footnotemark[$*$] $^{13}$CN in IC 342, and C$^{17}$O, C$^{34}$S, and H$^{13}$CO$^{+}$ in NGC 1068 are not detected. Optical depths of CS and HCO$^{+}$ in NGC 1068 are not calculated using the elemental isotopic ratios because C$^{34}$S and H$^{13}$CO$^{+}$ are not detected. The values of C$_{2}$H, CN, and $^{13}$CN are calculated as the sum of the optical depths of all components.
  }\hss}}
 \end{tabular}}
\end{table}

\clearpage

\begin{table*}
  \caption{Rotation temperatures and column densities of the observed molecules.}\label{}
  \begin{center}
    \begin{tabular}{lcccccccc}
      \hline
      & \multicolumn{2}{c}{NGC 1068} & & \multicolumn{2}{c}{NGC 253} & & \multicolumn{2}{c}{IC 342} \\
      \cline{2-3}
      \cline{5-6}
      \cline{8-9}
      Molecule & $T_{\rm rot}$ [K] & $N_{\rm mol}$ [cm$^{-2}$] & & $T_{\rm rot}$ [K] & $N_{\rm mol}$ [cm$^{-2}$] & & $T_{\rm rot}$ [K] & $N_{\rm mol}$ [cm$^{-2}$] \\
      \hline
cyclic-C$_{3}$H$_{2}$ & 10$\pm$5\footnotemark[$*$] & 2.0$^{+1.7}_{-1.0}$(14) & & 9.8$\pm$0.2 & 4.4$\pm$0.0(13) & & 10$\pm$5\footnotemark[$*$] & 6.9$^{+7.8}_{-4.3}$(13) \\

CH$_{3}$CCH & 10$\pm$5\footnotemark[$*$] & $<$9.1(13) & & 40.5$\pm$0.2 & 3.5$\pm$0.3(14) & & 39.6$^{+0.9}_{-0.8}$ & 4.1$\pm0.1$(14) \\

H$^{13}$CN & $<$7.3 & $>$4.7(13) & & 6.1$^{+1.1}_{-0.9}$ & 1.2$^{+0.2}_{-0.1}$(13) & & 10$\pm$5\footnotemark[$*$] & 1.5$^{+0.6}_{-0.4}$(13) \\

H$^{13}$CO$^{+}$ & 10$\pm$5\footnotemark[$*$] & $<$4.5(12) & & 6.7$^{+1.7}_{-1.1}$ & 2.5$\pm$0.3(12) & & 10$\pm$5\footnotemark[$*$] & 4.2$^{+2.1}_{-1.4}$(12) \\

SiO & 5.9$^{+3.2}_{-1.5}$ & 3.2$^{+1.2}_{-0.7}$(13) & & 7.2$^{+0.7}_{-0.5}$ & 1.1$\pm$0.1(13) & & 10$\pm$5\footnotemark[$*$] & 1.4$^{+0.5}_{-0.2}$(13) \\

C$_{2}$H & 10$\pm$5\footnotemark[$*$] & 2.6$\pm$1.0(15) & & 6.5$\pm$1.0 & 1.7$\pm$0.0(15) & & 10$\pm$5\footnotemark[$*$] & 5.7$^{+2.2}_{-2.1}$(14) \\

HNCO & 39.9$^{+9.5}_{-4.9}$ & 6.8$^{+0.6}_{-0.1}$(14) & & 19.1$\pm$2.0 & 1.1$^{+0.1}_{-0.0}$(14) & & 24.0$^{+1.1}_{-0.9}$ & 2.5$^{+0.1}_{-0.0}$(14) \\

HCN & 10.3$^{+0.2}_{-0.5}$ & 4.7$^{+0.0}_{-0.3}$(14) & & 13.9$^{+0.0}_{-0.2}$ & 2.0$^{+0.1}_{-0.0}$(14) & & 10.8$\pm$0.1 & 1.2$\pm$0.1(14) \\

HCO$^{+}$ & 8.3$\pm$0.3 & 2.1$\pm$0.1(14) & & 13.2$\pm$0.2 & 1.0$^{+0.3}_{-0.4}$(14) & & 6.2$\pm$0.3 & 7.5$^{+0.1}_{-0.0}$(13) \\

HNC & 9.5$^{+0.4}_{-0.5}$ & 1.7$\pm$0.1(14) & & 10$\pm$5\footnotemark[$*$] & 8.7$^{+2.6}_{-1.7}$(13) & & 10$\pm$5\footnotemark[$*$] & 7.6$^{+2.2}_{-1.4}$(13) \\
HC$_{3}$N & 13.4$^{+32.2}_{-5.0}$ & 1.7$^{+2.7}_{-0.9}$(14) & & 36.8$^{+2.6}_{-2.8}$ & 3.3$^{+0.3}_{-0.2}$(13) & & 33.8$^{+4.3}_{-3.7}$ & 2.6$\pm$0.3(13) \\

N$_{2}$H$^{+}$ & 10$\pm$5\footnotemark[$*$] & 5.8$^{+2.9}_{-1.8}$(13) & & 10$\pm$5\footnotemark[$*$] & 3.1$^{+1.0}_{-0.6}$(13) & & 10$\pm$5\footnotemark[$*$] & 3.5$^{+1.2}_{-0.8}$(13) \\

C$^{34}$S & $>$13.1 & $<$1.9(13) & & 15.1$^{+10.1}_{-4.3}$ & 2.7$^{+0.7}_{-0.1}$(13) & & 10.4$^{+14.8}_{-3.8}$ & 1.8$^{+0.5}_{-0.0}$(13) \\

CH$_{3}$OH & 10$\pm$5\footnotemark[$*$] & 1.7$^{+0.3}_{-0.1}$(13) & & 10$\pm$5\footnotemark[$*$] & 1.7$^{+0.4}_{-0.1}$(13) & & 10$\pm$5\footnotemark[$*$] & 1.9$^{+0.5}_{-0.1}$(13) \\

CS & 7.9$\pm$0.3 & 5.3$^{+0.4}_{-0.0}$(14) & & 12.1$\pm$0.1 & 2.3$^{+0.0}_{-0.2}$(14) & & 9.8$^{+0.6}_{-0.7}$ & 2.1$^{+0.2}_{-0.3}$(14) \\

SO & $<$13.0 & $>$1.7(14) & & 7.4$^{+0.4}_{-0.3}$ & 2.0$\pm$0.0(14) & & $<$5.1 & $>$8.7(13) \\

$^{13}$CN & 10$\pm$5\footnotemark[$*$] & 1.0$^{+0.3}_{-0.2}$(14) & & 10$\pm$5\footnotemark[$*$] & 1.2$^{+0.5}_{-0.3}$(13) & & 10$\pm$5\footnotemark[$*$] & $<$4.4(12) \\

C$^{18}$O & 4.0$\pm$0.3 & 5.3$^{+0.6}_{-0.5}$(16) & & 19.1$^{+2.7}_{-2.3}$ & 3.5$\pm$0.1(16) & & 7.6$^{+0.9}_{-0.8}$ & 2.9$^{+0.6}_{-0.5}$(16) \\

$^{13}$CO & 11.7$\pm$0.8 & 1.2$\pm$0.0(17) & & 32.2$^{+3.8}_{-3.5}$ & 1.4$\pm$0.1(17) & & 12.5$^{+0.5}_{-0.4}$ & 1.7$\pm$0.0(17) \\

CH$_{3}$CN & 10$\pm$5\footnotemark[$*$] & 3.4$^{+6.1}_{-0.5}$(13) & & 9.5$\pm$0.1 & 1.0$^{+0.1}_{-0.0}$(13) & & 10$\pm$5\footnotemark[$*$] & 9.6$^{+17.1}_{-1.2}$(12) \\

C$^{17}$O & 10$\pm$5\footnotemark[$*$] & $<$4.1(15) & & 26.5\footnotemark[$\dagger$] & 3.7\footnotemark[$\dagger$] & & 7.0$^{+3.5}_{-1.7}$ & 4.2$^{+1.4}_{-0.9}$(15) \\

CN & 4.9$\pm$0.2 & 3.0$\pm$0.2(15) & & 5.4$\pm$0.1 & 1.1$\pm$0.0(15) & & 3.6$\pm$0.1 & 8.1$^{+0.3}_{-0.1}$(14) \\

$^{12}$CO & 19.0$\pm$1.0 & 1.9$\pm$0.1(18) & & 42.8$^{+0.0}_{-10.3}$ & 1.3$^{+0.0}_{-0.2}$(18) & & 24.4$^{+0.2}_{-0.3}$ & 8.2$\pm$0.1(17) \\
            \hline
      \multicolumn{5}{@{}l@{}}{\hbox to 0pt{\parbox{138mm}{\footnotesize
       The expression a(b) represents $a\times 10^{b}$. The range of error is represented by the maximum and minimum values, which are calculated from the maximum and minimum slopes of the fitted lines in the rotation diagram. 
    \par\noindent
\footnotemark[$*$] We assumed $T_{\rm{rot}}$ = 10$\pm$5 K for molecules with only one transition available or non-detection in our observations.  
    \par\noindent
\footnotemark[$\dagger$] The error of C$^{17}$O in NGC 253 is not calculated because the fitted line can have a positive slope due to a large error in our observations (see the rotaion diagram in figure 10).
     }\hss}}
    \end{tabular}
  \end{center}
\end{table*}

\clearpage

\begin{table}
 \tbl{Important column density ratios.}{%
 \begin{tabular}{lccc}
  \hline
  Column density ratio & NGC 1068 & NGC 253 & IC 342 \\
  \hline
  HCN/HCO$^{+}$ & 2.2$^{+0.1}_{-0.2}$ & 2.0$^{+0.1}_{-0.8}$ & 1.6$^{+0.6}_{-0.1}$ \\
  CN/HCO$^{+}$ & 14.3$\pm{1.2}$ & 11.0$^{+3.3}_{-4.4}$ & 10.8$^{+0.4}_{-0.1}$ \\
  H$^{13}$CN/H$^{13}$CO$^{+}$ & $>$10.4 & 4.8$^{+1.0}_{-0.6}$ & 3.6$^{+2.3}_{-1.5}$ \\
  $^{13}$CN/H$^{13}$CO$^{+}$ & $>$22.2 & 4.8$^{+2.1}_{-1.3}$ & $<$1.0 \\
  CN/HCN & 6.4$^{+0.4}_{-0.0}$ & 5.5$^{+0.3}_{-0.0}$ & 6.8$\pm{0.6}$ \\
  $^{13}$CN/H$^{13}$CN & $<$2.1 & 1.0$^{+0.4}_{-0.3}$ & $<$0.3 \\
  CH$_{3}$CCH/C$^{18}$O & $<$1.7$\times$10$^{-3}$ & 10.0$\pm$0.1$\times$10$^{-3}$ & 14.1$\pm$0.3$\times$10$^{-3}$ \\
  \hline
 \end{tabular}}\label{}
\end{table}

\clearpage

\begin{figure*}
   \begin{center}
      \FigureFile(160mm,80mm){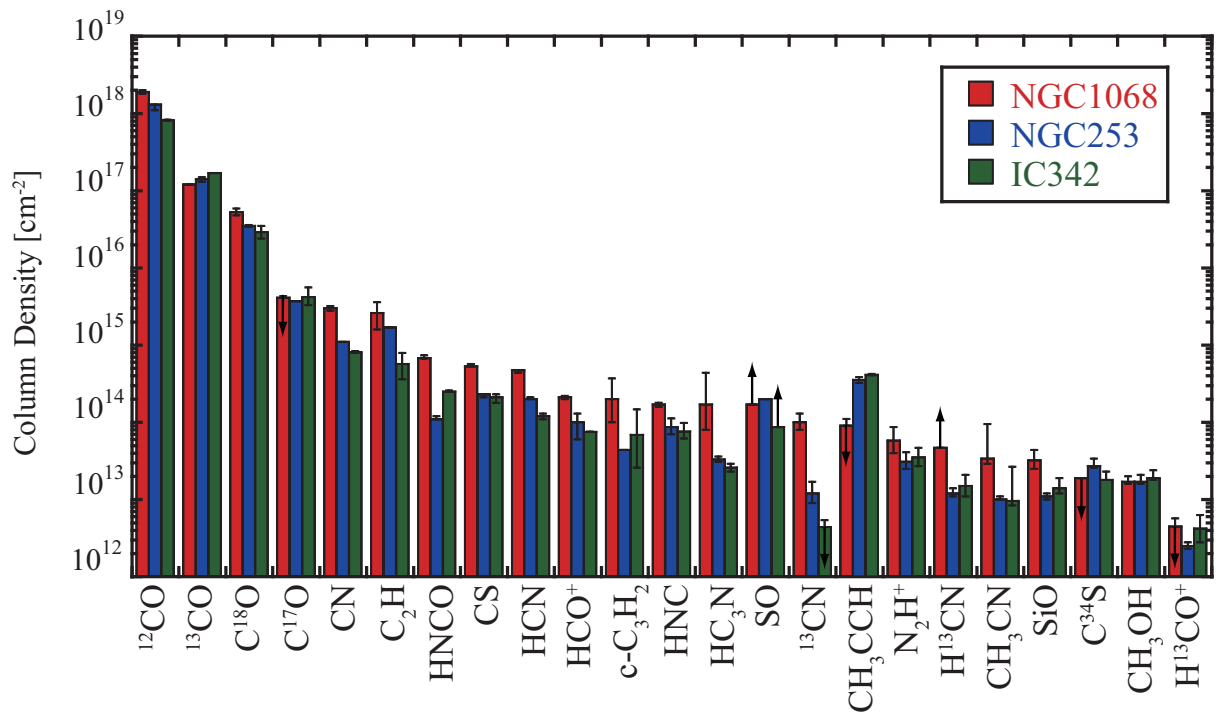}
   \end{center}
   \caption{Comparison of column densities of each molecule among the three galaxies. The bar graphs of red (left), blue (center), and green (right) are for NGC 1068, NGC 253, and IC 342, respectively. The order of the molecules is arranged in descending order from left to right based on the column density toward NGC 1068. Arrows indicate upper or lower limits (see also Table 3).}\label{somelabel}
\end{figure*}

\clearpage

\begin{figure}
   \begin{center}
      \FigureFile(80mm,40mm){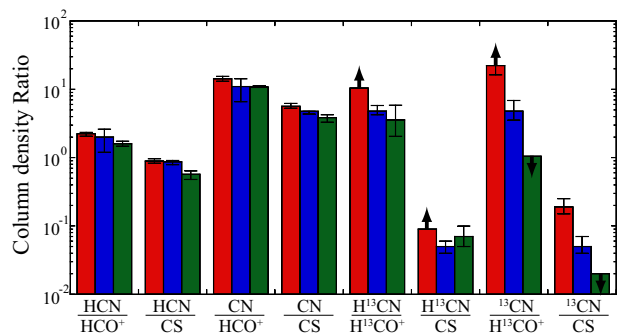}
   \end{center}
   \caption{The important column density ratios among the observed three galaxies. The numerical values are shown in Table 4. Arrows indicate upper or lower limits.}\label{somelabel}
\end{figure}

\clearpage

\begin{figure}
   \begin{center}
      \FigureFile(80mm,80mm){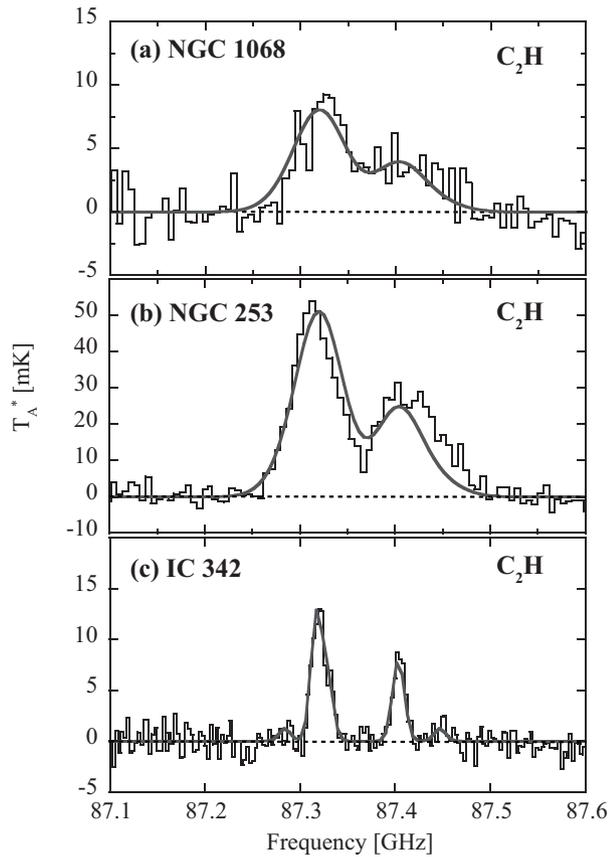}
   \end{center}
   \caption{C$_{2}$H $N$ = 1--0 lines toward (a) NGC 1068, (b) NGC 253, and (c) IC 342 obtained with a velocity resolution of 20 km s$^{-1}$ for NGC 1068 and NGC 253 and 10 km s$^{-1}$ for IC 342. The spectrum consists of 6 hyperfine components ({\it J} = 3/2--1/2; {\it F} = 1--1, {\it F} = 2--1, {\it F} = 1--0, and {\it J} = 1/2--1/2; {\it F} = 1--1, {\it F} = 0--1, {\it F} = 1--0). The gray curves overlaid on the profiles are the results of Gaussian least-squares fits.}\label{somelabel}
\end{figure}

\begin{figure}
   \begin{center}
      \FigureFile(80mm,80mm){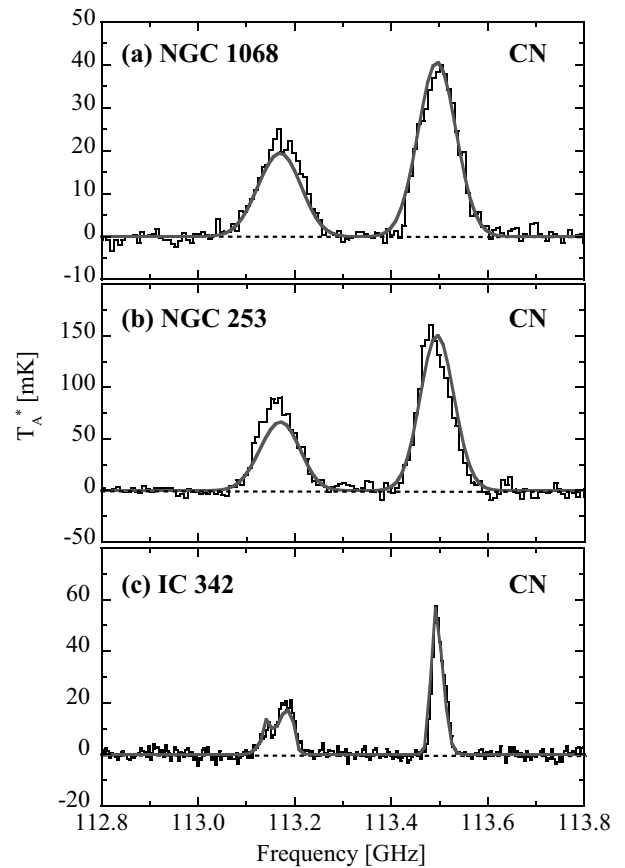}
   \end{center}
   \caption{CN $N$ = 1--0 lines toward (a) NGC 1068, (b) NGC 253, and (c) IC 342. The spectrum consists of 9 hyperfine components ({\it J} = 3/2--1/2; {\it F} = 3/2--1/2, {\it F} = 5/2--3/2, {\it F} = 1/2--1/2, {\it F} = 3/2--3/2, {\it F} = 1/2--3/2) and {\it J} = 1/2--1/2; {\it F} = 1/2--1/2, {\it F} = 1/2--3/2, {\it F} = 3/2--1/2 , {\it F} = 3/2--3/2. See also the caption of figure 3.}\label{somelabel}
\end{figure}

\begin{figure}
   \begin{center}
      \FigureFile(80mm,80mm){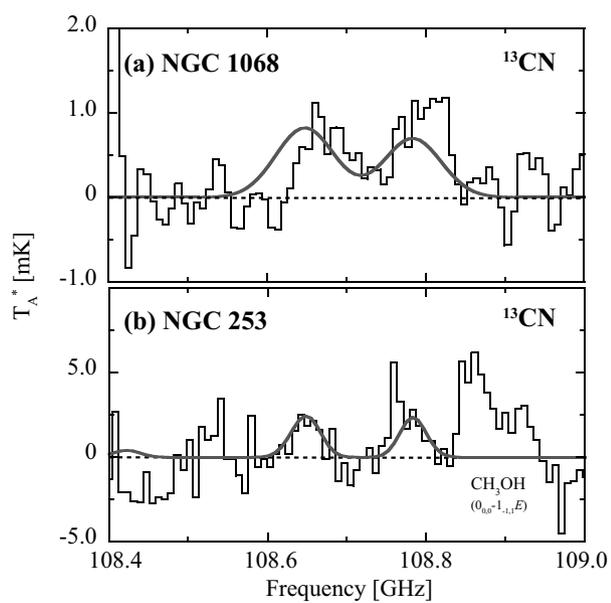}
   \end{center}
   \caption{$^{13}$CN $N$ = 1--0 lines toward (a) NGC 1068 and (b) NGC 253. The spectrum consists of 14 hyperfine components. The reason for the possible fitting offset in NGC 1068 may have resulted from a low signal to noise ratio. See also the caption of figure 3.}\label{somelabel}
\end{figure}

\clearpage

\begin{figure*}
   \begin{center}
      \FigureFile(160mm,80mm){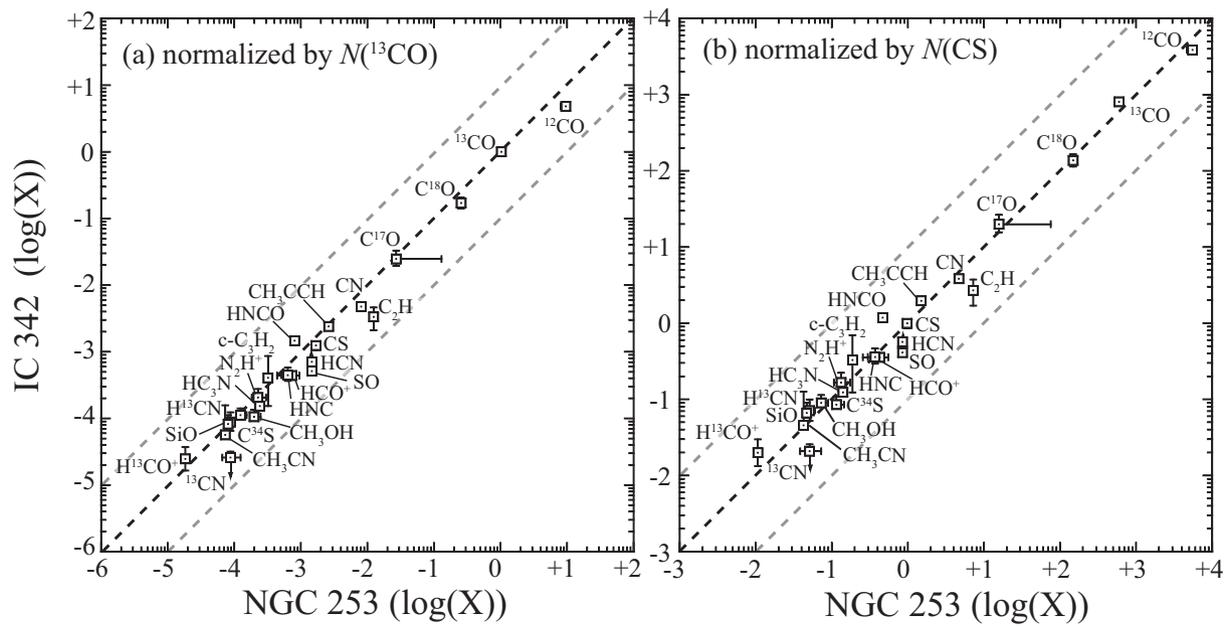}
   \end{center}
   \caption{Plots of the fractional molecular abundances relative to $^{13}$CO (a) and CS (b) between the two starburst galaxies, NGC 253 and IC 342. The bold dashed line represents equal abundances between NGC 253 and IC 342, and thin dashed lines represent an order of magnitude higher or lower than the equal abundance. For details of the normalization for the calculation of the fractional abundances, see section 4.1.}\label{somelabel}
\end{figure*}

\begin{figure*}
   \begin{center}
      \FigureFile(160mm,160mm){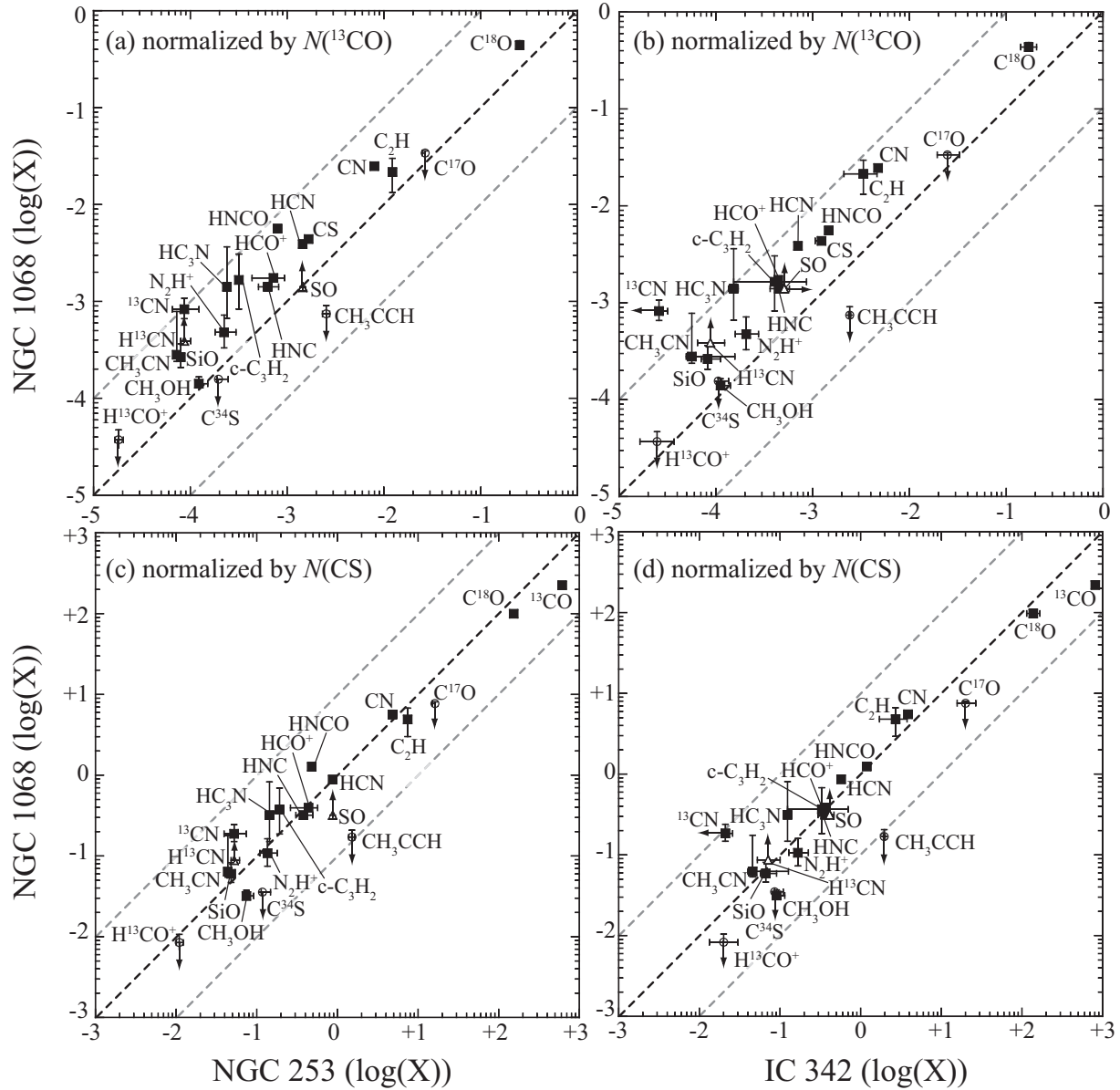}
   \end{center}
   \caption{Plots of the fractional molecular abundances relative to $^{13}$CO ((a) and (b)) and CS ((c) and (d)) between NGC 253 and NGC 1068, and between IC 342 and NGC 1068. These results represent the abundance correlation between the galaxy with an AGN and the two starburst galaxies. Plots above the bold dashed line represent enhanced molecules and plots below represent deficient ones in the galaxy hosting the AGN. For details of the normalization for the calculation of the fractional abundances, see section 4.1.}\label{somelabel}
\end{figure*}

\clearpage

\begin{figure*}
   \begin{center}
      \FigureFile(160mm,160mm){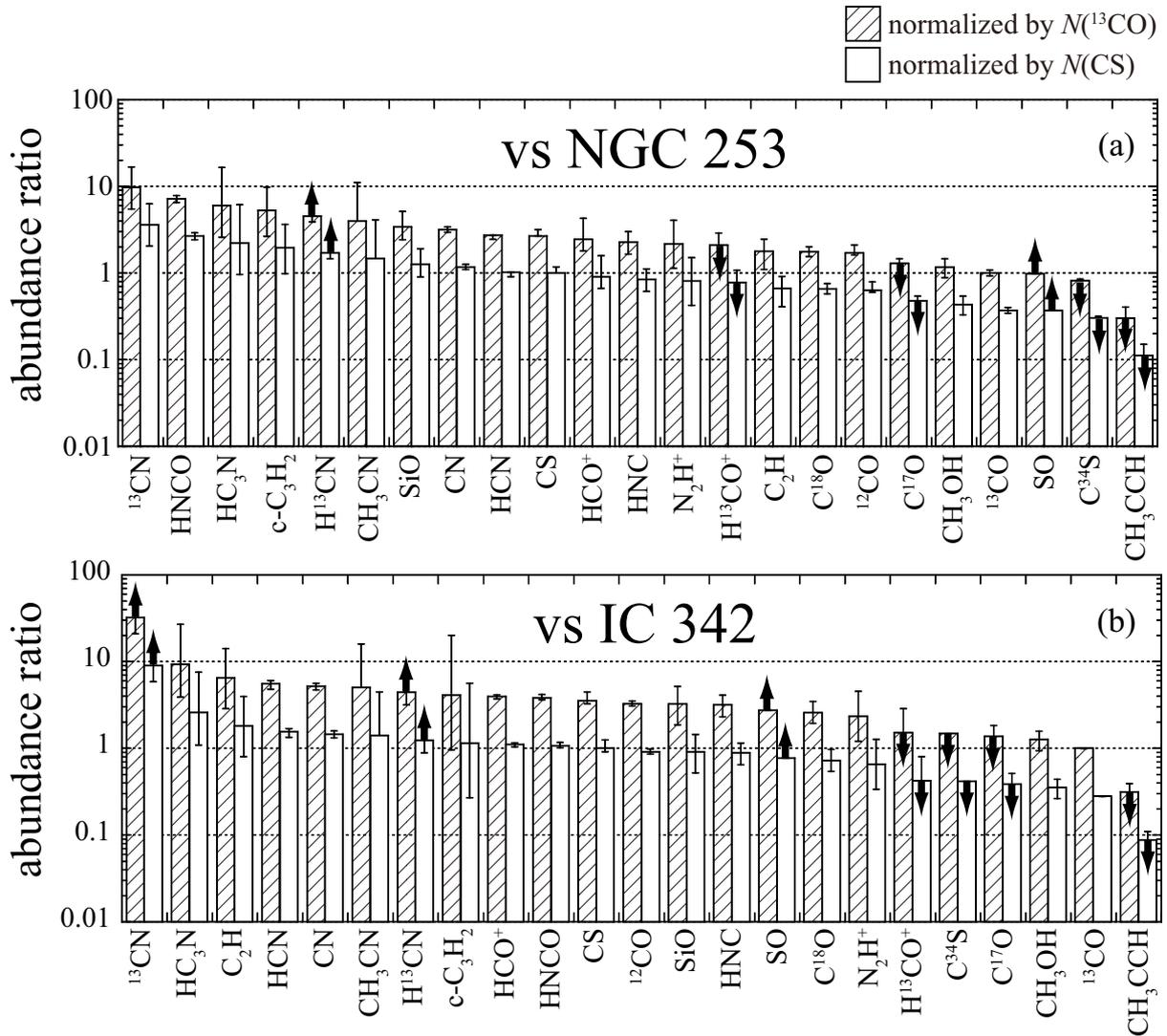}
   \end{center}
   \caption{Fractional abundances relative to $^{13}$CO and CS in NGC 1068 compared with (a) NGC 253 and (b) IC 342. Ratios over unity represents enhancement of molecular abundances in NGC 1068. Arrows indicate upper or lower limits. For details of the normalization for the calculation of the fractional abundances, see section 4.1. The abscissa consists of the 23 molecules detected in at least one of the three galaxies.}\label{somelabel}
\end{figure*}

\clearpage

\appendix
\section{Rotation diagram of each molecule}

\begin{figure*}
   \begin{center}
      \FigureFile(160mm,300mm){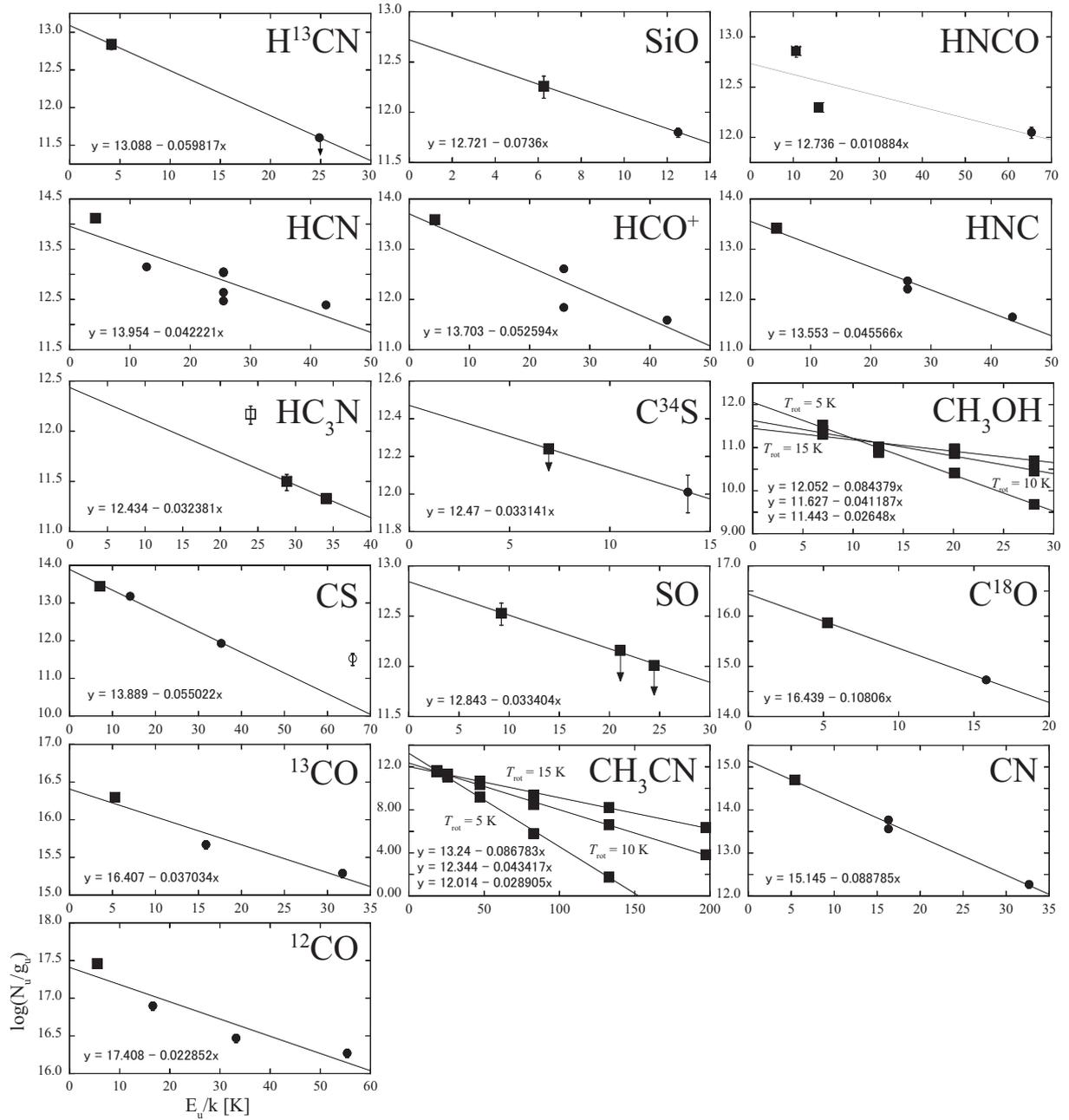}
   \end{center}
   \caption{Rotational diagram of each detected molecule in NGC 1068. The filled square symbols represent the results of this work, and the filled circle symbols represent the results of previous studies. The open square and open circle symbols indicate marginally detected molecules, and we do not use these data for the fitting in the rotation diagrams. For details of each molecule and the line parameters, see section 3.1 and Appendix 2.}\label{somelabel}
\end{figure*}

\clearpage

\begin{figure*}
   \begin{center}
      \FigureFile(160mm,300mm){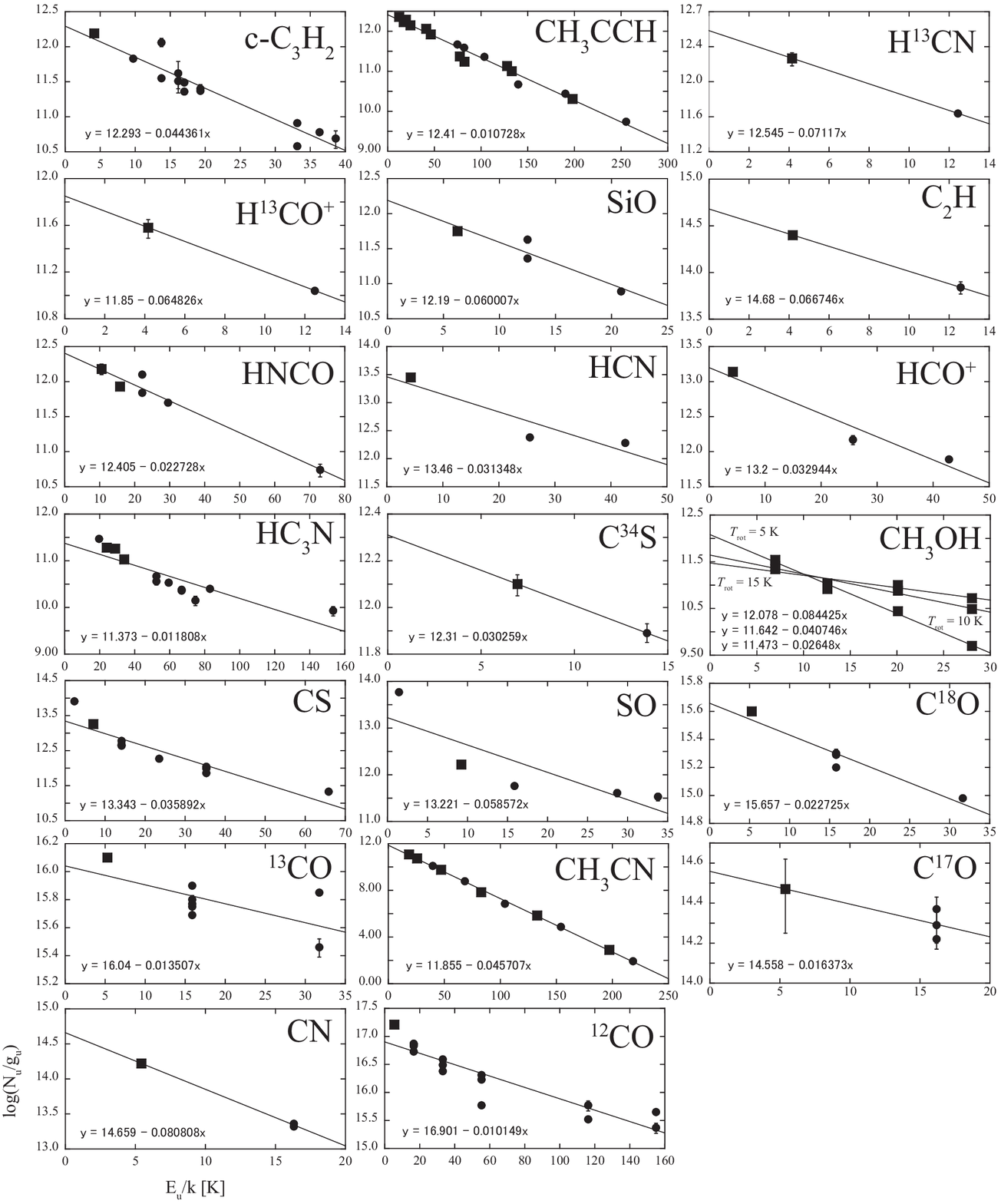}
   \end{center}
   \caption{Rotational diagram of each detected molecule in NGC 253. For details, see the caption of figure 9. We calculated only the $K$ = 0 to 5 components of CH$_{3}$CN and CH$_{3}$CCH. Higher values of the $K$-ladder (K $>$ 6) were not taken into account, because the $K$ = 5 component has a contribution of less than 1 \% to the total integrated intensity.}\label{somelabel}
\end{figure*}

\clearpage

\begin{figure*}
   \begin{center}
      \FigureFile(160mm,300mm){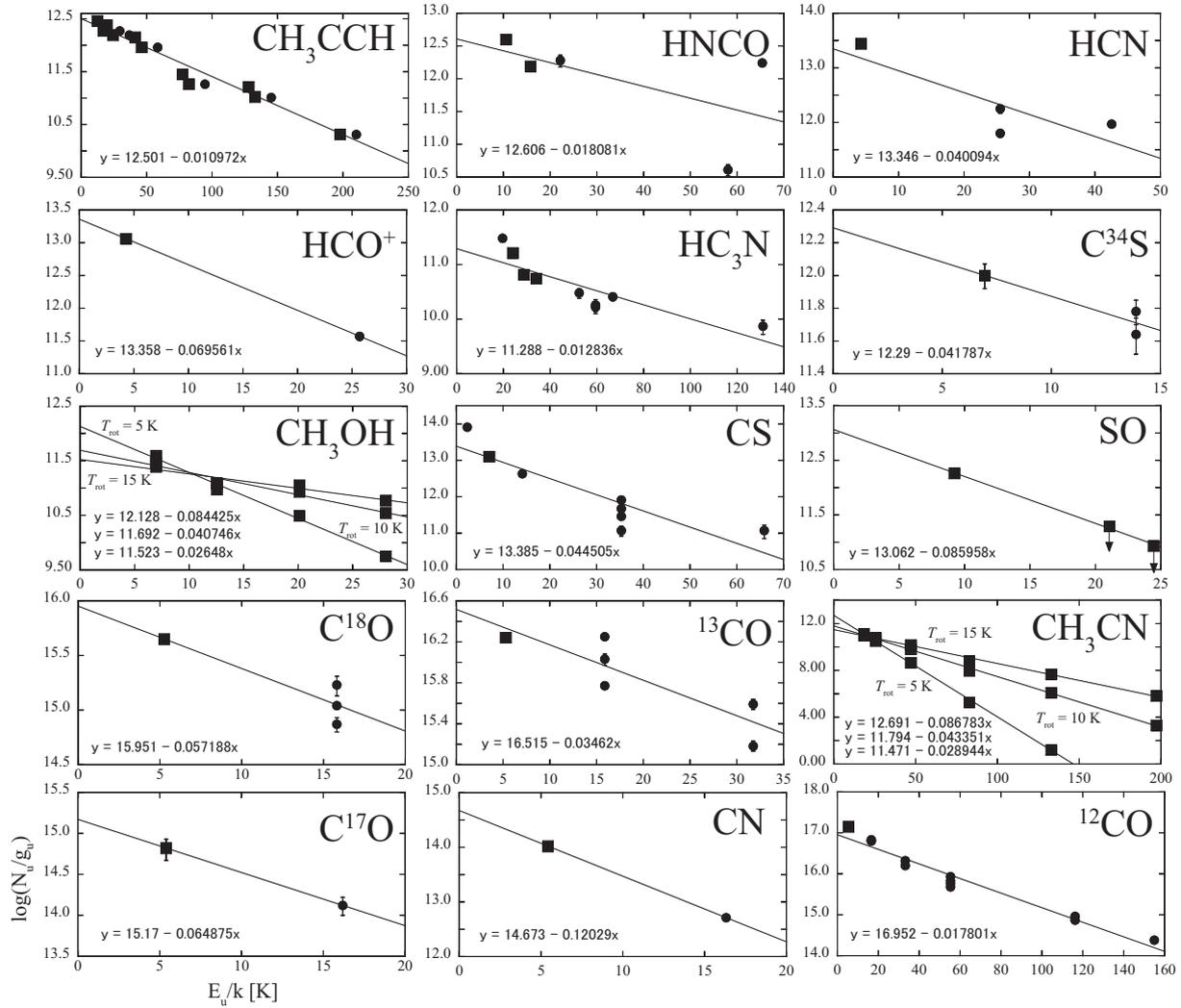}
   \end{center}
   \caption{Rotational diagram of each detected molecule in IC 342. For details, see the caption of figure 9. We calculated only the $K$ = 0 to 5 components of CH$_{3}$CCH. Higher values of the $K$-ladder (K $>$ 6) were not taken into account, because the $K$ = 5 component having a contribution of less than 1 \% to the total integrated intensity.}\label{somelabel}
\end{figure*}

\clearpage

\section{Parameters of each molecular line for the rotation diagrams}

\begin{longtable}{lccccccc}
  \caption{NGC 1068.}
      \hline
      Molecule & Transition & $E\rm{u}/\it{k}$ & Telescope & $\theta_{\rm b}$ & $T_{\rm mb}$ & $\int$$T_{\rm mb}dv$ & Reference \\
       & & [K] & & [$^{\prime\prime}$] & [mK] & [K km s$^{-1}$] & \\
      \hline
      \endhead
      \hline
      \endfoot
      \hline
      \multicolumn{8}{l}{\hbox to 0pt{\parbox{125mm}{\footnotesize
       References. (1) Takano et al. in prep.; (2) Wang et al. 2014; (3) Usero et al. 2004; (4) Mart\'{i}n et al. 2009; (5) Krips et al. 2008; (6) Paglione et al. 1997; (7) P\'{e}rez-Beaupuits et al. 2007; (8) Kamenetzky et al. 2011; (9) P\'{e}rez-Beaupuits et al. 2009; (10) Mauersberger et al. 1989; (11) Bayet et al. 2009; (12) Israel 2009.
      }}}
      \endlastfoot
      H$^{13}$CN & 1--0 & 4.15 & NRO 45-m & 18.7 & 7$\pm$1 & 1.4$\pm$0.2 & 1 \\
      H$^{13}$CN & 3--2 & 24.88 & APEX 12-m & 23.5 & & $<$0.46 & 2 \\
      &&&&&&&\\
      SiO & 2--1 & 6.26 & NRO 45-m & 18.7 & 5$\pm$1 & 0.8$\pm$0.2 & 1 \\
      SiO & 3--2 & 12.51 & IRAM 30-m & 19 & 3.0 & 0.60$\pm$0.06 & 3 \\
      &&&&&&&\\
      HNCO & 4$_{0,4}$--3$_{0,3}$ & 10.56 & NRO 45-m & 19.1 & 6$\pm$3 & 1.6$\pm$0.2 & 1 \\
      HNCO & 5$_{0,5}$--4$_{0,4}$ & 15.84 & NRO 45-m & 16.5 & 4$\pm$1 & 0.9$\pm$0.1 & 1 \\
      HNCO & 6$_{1,6}$--5$_{1,5}$ & 65.40 & IRAM 30-m & 19.0 & 2.5$\pm$0.6 & 0.56$\pm$0.07 & 3 \\
      &&&&&&&\\
      HCN & 1--0 & 4.26 & NRO 45-m & 19.1 & 108$\pm$2 & 26.4$\pm$0.6 & 1 \\
      HCN & 2--1 & 12.77 & IRAM 30-m & 14 & & 20.0$\pm$0.4 & 5 \\
      HCN & 3--2 & 25.54 & CSO 10-m & 27 & 32$\pm$7 & 4.0$\pm$0.6 & 6 \\
      HCN & 3--2 & 25.54 & JCMT 15-m & 18 & 81.2$\pm$4.0 & 22.7$\pm$2.5 & 7 \\
      HCN & 3--2 & 25.54 & IRAM 30-m & 9.5 & & 19.0$\pm$0.6 & 5 \\
      HCN & 3--2 & 25.54 & CSO 10-m & 29 & & 8.5$\pm$0.3 & 8 \\
      HCN & 4--3 & 42.57 & JCMT 15-m & 14 & 65.9$\pm$3.8 & 13.9$\pm$1.6 & 9 \\
      &&&&&&&\\
      HCO$^{+}$ & 1--0 & 4.28 & NRO 45-m & 19.1 & 55$\pm$2 & 13.3$\pm$0.7 & 1 \\
      HCO$^{+}$ & 3--2 & 25.70 & IRAM 30-m & 9.5 & & 7.6$\pm$0.8 & 5 \\
      HCO$^{+}$ & 3--2 & 25.70 & CSO 10-m & 29 & & 5.58$\pm$0.35 & 8 \\
      HCO$^{+}$ & 4--3 & 42.84 & JCMT 15-m & 14 & 23.7$\pm$2.2 & 3.8$\pm$0.5 & 9 \\
      &&&&&&&\\
      HNC & 1--0 & 4.35 & NRO 45-m & 19.1 & 28$\pm$2 & 5.6$\pm$0.4 & 1 \\
      HNC & 3--2 & 26.13 & JCMT 15-m & 18 & 21.5$\pm$3 & 3.5$\pm$0.6 & 7 \\
      HNC & 3--2 & 26.13 & CSO 10-m & 28 & & 2.15$\pm$0.21 & 8 \\
      HNC & 4--3 & 43.55 & JCMT 15-m & 14 & 10.6$\pm$1.7 & 2.7$\pm$0.3 & 9 \\
      &&&&&&&\\
      HC$_{3}$N & 10--9 & 24.04 & NRO 45-m & 19.1 & 16$\pm$4 & 4.7$\pm$1.0 & 1 \\
      HC$_{3}$N & 11--10 & 28.84 & NRO 45-m & 16.5 & 9$\pm$2 & 1.6$\pm$0.3 & 1 \\
      HC$_{3}$N & 12--11 & 34.09 & NRO 45-m & 16.5 & 5$\pm$1 & 1.3$\pm$0.1 & 1 \\
      &&&&&&&\\
      C$^{34}$S & 2--1 & 6.95 & NRO 45-m & 19 & & $<$0.33 & 1 \\
      C$^{34}$S & 3--2 & 13.89 & IRAM 30-m & 17 & 2.2$\pm$1.0 & 0.53$\pm$0.12 & 4 \\
      &&&&&&&\\
      CH$_{3}$OH & 2$_{K}$--1$_{K}$ & 6.97 & NRO 45-m & 19 & 5$\pm$1 & 1.1$\pm$0.2 & 1 \\
      &&&&&&&\\
      CS & 2--1 & 7.06 & NRO 45-m & 19 & 22$\pm$1 & 5.4$\pm$0.2 & 1 \\
      CS & 3--2 & 14.12 & IRAM 30-m & 16 & 30 & 9.1$\pm$1.5 & 10 \\
      CS & 5--4 & 35.30 & IRAM 30-m & 10 & 17.6$\pm$3.0 & 3.3$\pm$0.3 & 4 \\
      CS & 7--6 & 65.88 & JCMT 15-m & 14 & 4.8$\pm$8.1 & 1.4$\pm$0.5 & 11 \\
      &&&&&&&\\
      SO & 3$_{2}$--2$_{1}$ & 9.23 & NRO 45-m & 16.5 & 4$\pm$1 & 0.8$\pm$0.2 & 1 \\
      SO & 2$_{3}$--1$_{2}$ & 21.07 & NRO 45-m & 16.5 & & $<$0.25 & 1 \\
      SO & 4$_{5}$--4$_{4}$ & 24.45 & NRO 45-m & 16.5 & & $<$0.4 & 1 \\
      &&&&&&&\\
      C$^{18}$O & 1--0 & 5.27 & NRO 45-m & 16.5 & 11$\pm$1 & 3.3$\pm$0.2 & 1 \\
      C$^{18}$O & 2--1 & 15.82 & IRAM 30-m & 11 & 9.0$\pm$1.9 & 2.0$\pm$0.2 & 4 \\
      &&&&&&&\\
      $^{13}$CO & 1--0 & 5.29 & NRO 45-m & 16.5 & 36$\pm$1 & 8.9$\pm$0.2 & 1 \\
      $^{13}$CO & 2--1 & 15.88 & IRAM 30-m & 12 & 56 & 15.0$\pm$1.8 & 12 \\
      $^{13}$CO & 3--2 & 31.76 & JCMT 15-m & 14 & 43 & 10.7$\pm$1.3 & 12 \\
      &&&&&&&\\
      CH$_{3}$CN & 6$_{K}$--5$_{K}$ & 18.56 & NRO 45-m & 16.5 & 3$\pm$1 & 1.0$\pm$0.2 & 1 \\
      &&&&&&&\\
      CN & 1--0 & 5.43 & NRO 45-m & 15.2 & 110$\pm$2 & 46.1$\pm$0.5 & 1 \\
      CN & 2--1 & 16.32 & SEST 15-m & 23.0 & 20.4$\pm$1.8 & 6.1$\pm$0.7 & 7 \\
      CN & 2--1 & 16.32 & JCMT 15-m & 21.0 & 31.9$\pm$2.5 & 11.7$\pm$1.4 & 9 \\
      CN & 3--2 & 32.66 & JCMT 15-m & 15.0 & 7.9$\pm$1.4 & 1.6$\pm$0.3 & 9 \\
      &&&&&&&\\
      $^{12}$CO & 1--0 & 5.54 & NRO 45-m & 15.2 & 638$\pm$16 & 155$\pm$1 & 1 \\
      $^{12}$CO & 2--1 & 16.61 & IRAM 30-m & 12 & 1080 & 266$\pm$32 & 12 \\
      $^{12}$CO & 3--2 & 33.22 & JCMT 15-m & 14 & 630 & 166$\pm$20 & 12 \\
      $^{12}$CO & 4--3 & 55.37 & JCMT 15-m & 11 & 1060 & 290$\pm$35 & 12 \\
\end{longtable}

\begin{longtable}{lccccccc}
  \caption{NGC 253.}
      \hline
      Molecule & Transition & $E\rm{u}/\it{k}$ & Telescope & $\theta_{\rm b}$ & $T_{\rm mb}$ & $\int$$T_{\rm mb}dv$ & Reference \\
       & & [K] & & [$^{\prime\prime}$] & [mK] & [K km s$^{-1}$] & \\
      \hline
      \endhead
      \hline
      \endfoot
      \hline
      \multicolumn{8}{l}{\hbox to 0pt{\parbox{135mm}{\footnotesize
       References. (1) Takano et al. in prep.; (2) Mart\'{i}n et al. 2006; (3) Oike et al. 2004; (4) Aladro et al. 2011a; (5) Mauersberger et al. 1991; (6) Nguyen-Q-Rieu et al. 1991; (7) Paglione et al. 1997; (8) Jackson et al. 1995; (9) Nguyen-Q-Rieu et al. 1992; (10) Mauersberger et al. 1990; (11) Mart\'{i}n et al. 2005; (12) Paglione et al. 1995; (13) Mauersberger \& Henkel 1989; (14) Mauersberger et al. 2003; (15) Bayet et al. 2009; (16) Takano et al. 1995; (17) Sage et al. 1991; (18) Harrison et al. 1999; (19) Wall et al. 1991; (20) Henkel et al. 1988; (21) Mauersberger et al. 1996; (22) Israel et al. 1995; (23) G$\ddot{\rm u}$sten et al. 1993; (24) Israel et al. 2002; (25) Harris et al. 1991; (26) Bayet et al. 2004; (27) Bradford et al. 2003. 
      }}}
      \endlastfoot
      cyclic-C$_{3}$H$_{2}$ & 2$_{1,2}$--1$_{0,1}$ & 4.15 & NRO 45-m & 18.7 & 33$\pm$2 & 5.1$\pm$0.4 & 1 \\
      cyclic-C$_{3}$H$_{2}$ & 2$_{2,0}$--1$_{1,1}$ & 9.72 & IRAM 30-m & 16.7 & 3.6 & 0.6 & 2 \\
      cyclic-C$_{3}$H$_{2}$ & 3$_{1,2}$--3$_{0,3}$ & 13.76 & IRAM 30-m & 29 & 12$\pm$4 & 1.44$\pm$0.20 & 3 \\
      cyclic-C$_{3}$H$_{2}$ & 3$_{1,2}$--2$_{2,1}$ & 13.76 & IRAM 30-m & 17.3 & 11 & 1.75 & 2 \\
      cyclic-C$_{3}$H$_{2}$ & 3$_{2,2}$--2$_{1,1}$ & 16.15 & IRAM 30-m & 16 & 8$\pm$5 & 1.05$\pm$0.50 & 3 \\
      cyclic-C$_{3}$H$_{2}$ & 3$_{2,2}$--2$_{1,1}$ & 16.15 & IRAM 30-m & 16.1 & 6.6 & 0.82$\pm$0.18 & 2 \\
      cyclic-C$_{3}$H$_{2}$ & 4$_{1,4}$--3$_{0,3}$ & 17.02 & IRAM 30-m & 16 & 39$\pm$4 & 4.76$\pm$0.20 & 3 \\
      cyclic-C$_{3}$H$_{2}$ & 4$_{1,4}$--3$_{0,3}$ & 17.02 & IRAM 30-m & 16.6 & 20 & 3.38$\pm$0.11 & 2 \\
      cyclic-C$_{3}$H$_{2}$ & 4$_{0,4}$--3$_{1,3}$ & 19.32 & IRAM 30-m & 16 & & 1.28$\pm$0.20 & 3 \\
      cyclic-C$_{3}$H$_{2}$ & 4$_{0,4}$--3$_{1,3}$ & 19.32 & IRAM 30-m & 16.6 & 7.0 & 1.15 & 2 \\
      cyclic-C$_{3}$H$_{2}$ & 5$_{1,4}$--5$_{0,5}$ & 33.15 & IRAM 30-m & 16.6 & 1.0 & 0.165 & 2 \\
      cyclic-C$_{3}$H$_{2}$ & 5$_{1,4}$--4$_{2,3}$ & 33.15 & IRAM 30-m & 11 & & 2.24$\pm$0.20 & 3 \\
      cyclic-C$_{3}$H$_{2}$ & 6$_{1,6}$--5$_{0,5}$ & 36.34 & IRAM 30-m & 11 & 30$\pm$4 & 2.64$\pm$0.20 & 3 \\
      cyclic-C$_{3}$H$_{2}$ & 6$_{0,6}$--5$_{1,5}$ & 38.64 & IRAM 30-m & 11 & & 0.71$\pm$0.20 & 3 \\
      &&&&&&&\\
      CH$_{3}$CCH & 5$_{K}$--4$_{K}$ & 12.31 & NRO 45-m & 18.7 & 10$\pm$2 & 2.3$\pm$0.2 & 1 \\
      CH$_{3}$CCH & 6$_{K}$--5$_{K}$ & 17.24 & NRO 45-m & 16.5 & 22$\pm$3 & 2.8$\pm$0.5 & 1 \\
      CH$_{3}$CCH & 13$_{K}$--12$_{K}$ & 74.70 & JCMT 15-m & 21.7 & 17.3$\pm$3.8 & 2.9$\pm$0.3 & 4 \\
      &&&&&&&\\
      H$^{13}$CN & 1--0 & 4.15 & NRO 45-m & 18.7 & 24$\pm$2 & 4.4$\pm$0.7 & 1 \\
      H$^{13}$CN & 2--1 & 12.44 & IRAM 30-m & 14.3 & 40 & 5.6$\pm$0.3 & 2 \\
      &&&&&&&\\
      H$^{13}$CO$^{+}$ & 1--0 & 4.17 & NRO 45-m & 18.7 & 12$\pm$4 & 1.6$\pm$0.3 & 1 \\
      H$^{13}$CO$^{+}$ & 2--1 & 12.50 & IRAM 30-m & 14.2 & 23 & 2.31$\pm$0.13 & 2 \\
      &&&&&&&\\
      SiO & 2--1 & 6.26 & NRO 45-m & 18.7 & 17$\pm$2 & 3.0$\pm$0.3 & 1 \\
      SiO & 3--2 & 12.51 & IRAM 30-m & 16 & 22 & 5.8$\pm$0.6 & 5 \\
      SiO & 3--2 & 12.51 & IRAM 30-m & 18.9 & 18 & 2.75$\pm$0.07 & 2 \\
      SiO & 4--3 & 20.86 & IRAM 30-m & 14.2 & 13 & 2.07$\pm$0.16 & 2 \\
      &&&&&&&\\
      C$_{2}$H & 1--0 & 4.19 & NRO 45-m & 18.4 & 122$\pm$3 & 43.2$\pm$1.5 & 1 \\
      C$_{2}$H & 2--1 & 12.58 & IRAM 30-m & 14.1 & 88 & 57.0$\pm$8.5 & 2 \\
      &&&&&&&\\
      HNCO & 4$_{0,4}$--3$_{0,3}$ & 10.56 & NRO 45-m & 19.1 & 23$\pm$4 & 4.2$\pm$0.7 & 1 \\
      HNCO & 5$_{0,5}$--4$_{0,4}$ & 15.84 & NRO 45-m & 16.5 & 25$\pm$3 & 4.2$\pm$0.3 & 1 \\
      HNCO & 6$_{0,6}$--5$_{0,5}$ & 22.17 & IRAM 30-m & 17 & 65$\pm$7 & 8.7$\pm$0.7 & 6 \\
      HNCO & 6$_{0,6}$--5$_{0,5}$ & 22.17 & IRAM 30-m & 18.7 & 32 & 4.40$\pm$0.23 & 2 \\
      HNCO & 7$_{0,7}$--6$_{0,6}$ & 29.56 & IRAM 30-m & 16.3 & 34 & 4.90$\pm$0.15 & 2 \\
      HNCO & 7$_{1,6}$--6$_{1,5}$ & 72.98 & IRAM 30-m & 16.2 & 6.4 & 0.54$\pm$0.11 & 2 \\
      &&&&&&&\\
      HCN & 1--0 & 4.26 & NRO 45-m & 19.1 & 341$\pm$7 & 70.1$\pm$0.6 & 1 \\
      HCN & 3--2 & 25.54 & CSO 10-m & 27 & 264$\pm$26 & 36.3$\pm$3.1 & 7 \\
      HCN & 4--3 & 42.57 & CSO 10-m & 20 & 360$\pm$70 & 71.4 & 8 \\
      &&&&&&&\\
      HCO$^{+}$ & 1--0 & 4.28 & NRO 45-m & 19.1 & 279$\pm$6 & 58.6$\pm$0.8 & 1 \\
      HCO$^{+}$ & 3--2 & 25.70 & IRAM 30-m & 10 & & 87.0$\pm$13.0 & 9 \\
      HCO$^{+}$ & 4--3 & 42.84 & CSO 10-m & 20 & 280$\pm$40 & 50.8 & 8 \\
      &&&&&&&\\
      HC$_{3}$N & 9--8 & 19.67 & IRAM 30-m & 29 & 86 & 5.8$\pm$0.6 & 10 \\
      HC$_{3}$N & 10--9 & 24.04 & NRO 45-m & 19.1 & 32$\pm$3 & 7.6$\pm$0.8 & 1 \\
      HC$_{3}$N & 11--10 & 28.84 & NRO 45-m & 16.5 & 55$\pm$3 & 9.9$\pm$0.2 & 1 \\
      HC$_{3}$N & 12--11 & 34.09 & NRO 45-m & 16.5 & 42$\pm$3 & 7.0$\pm$0.3 & 1 \\
      HC$_{3}$N & 15--14 & 52.44 & IRAM 30-m & 17 & 54 & 3.6$\pm$0.6 & 10 \\
      HC$_{3}$N & 15--14 & 52.44 & IRAM 30-m & 18.2 & 29 & 4.41$\pm$0.33 & 2 \\
      HC$_{3}$N & 16--15 & 59.43 & IRAM 30-m & 17.2 & 26 & 3.80 & 2 \\
      HC$_{3}$N & 17--16 & 66.86 & IRAM 30-m & 15 & 50 & 3.4$\pm$0.5 & 10 \\
      HC$_{3}$N & 17--16 & 66.86 & IRAM 30-m & 16.2 & 21 & 3.04$\pm$0.16 & 2 \\
      HC$_{3}$N & 18--17 & 74.73 & IRAM 30-m & 15.2 & 20 & 2.20$\pm$0.5 & 2 \\
      HC$_{3}$N & 19--18 & 83.03 & IRAM 30-m & 14.2 & 35 & 4.60$\pm$0.57 & 2 \\
      HC$_{3}$N & 26--25 & 153.38 & IRAM 30-m & 12 & 47 & 3.2$\pm$0.7 & 10 \\
      &&&&&&&\\
      C$^{34}$S & 2--1 & 6.95 & NRO 45-m & 19 & 17$\pm$2 & 2.9$\pm$0.3 & 1 \\
      C$^{34}$S & 3--2 & 13.89 & IRAM 30-m & 16.8 & 27.1$\pm$1.4 & 4.5$\pm$0.4 & 11 \\
      &&&&&&&\\
      CH$_{3}$OH & 2$_{K}$--1$_{K}$ & 6.97 & NRO 45-m & 19 & 70$\pm$3 & 14.5$\pm$0.3 & 1 \\
      &&&&&&&\\
      CS & 1--0 & 2.35 & NRO 45-m & 36 & 97$\pm$17 & 21.8$\pm$1.0 & 12 \\
      CS & 2--1 & 7.06 & NRO 45-m & 19 & 208$\pm$5 & 42.8$\pm$0.4 & 1 \\
      CS & 3--2 & 14.12 & IRAM 30-m & 16 & 159 & 27.8$\pm$2.7 & 13 \\
      CS & 3--2 & 14.12 & IRAM 30-m & 16 & 240 & 37$\pm$1 & 5 \\
      CS & 3--2 & 14.12 & IRAM 30-m & 17.1 & 111.2$\pm$4.1 & 25.6$\pm$0.3 & 11 \\
      CS & 4--3 & 23.53 & IRAM 30-m & 11.7 & 108.2$\pm$7.0 & 24.9$\pm$0.6 & 11 \\
      CS & 5--4 & 35.30 & IRAM 30-m & 11 & 147 & 23.5$\pm$2.3 & 13 \\
      CS & 5--4 & 35.30 & HHT 10-m & 32 & 45 & 7.7$\pm$0.9 & 14 \\
      CS & 5--4 & 35.30 & SEST 15-m & 21 & 46.7$\pm$10.8 & 9.8$\pm$0.4 & 11 \\
      CS & 7--6 & 65.88 & JCMT 15-m & 14.0 & 67.1$\pm$5.1 & 7.9$\pm$0.3 & 15 \\
      &&&&&&&\\
      SO & 1$_{0}$--0$_{1}$ & 1.44 & NRO 45-m & 56 & 12.8$\pm$2.6 & 2.8 & 16 \\
      SO & 3$_{2}$--2$_{1}$ & 9.23 & NRO 45-m & 16.5 & 25$\pm$2 & 4.2$\pm$0.3 & 1 \\
      SO & 4$_{3}$--3$_{2}$ & 15.87 & IRAM 30-m & 20.9 & 17.8$\pm$2.1 & 2.5$\pm$0.3 & 11 \\
      SO & 3$_{4}$--2$_{3}$ & 28.71 & IRAM 30-m & 17.7 & 11.4$\pm$2.9 & 1.7$\pm$0.3 & 11 \\
      SO & 4$_{4}$--3$_{3}$ & 33.80 & IRAM 30-m & 14.3 & 14 & 2.2$\pm$0.4 & 2 \\
      &&&&&&&\\
      C$^{18}$O & 1--0 & 5.27 & NRO 45-m & 16.5 & 90$\pm$4 & 19.1$\pm$0.6 & 1 \\
      C$^{18}$O & 2--1 & 15.82 & IRAM 30-m & 13 & & 44.8$\pm$3.0 & 17 \\
      C$^{18}$O & 2--1 & 15.82 & IRAM 30-m & 12 & & 37.7$\pm$3.8 & 18 \\
      C$^{18}$O & 2--1 & 15.82 & JCMT 15-m & 23 & & 27$\pm$3 & 17 \\
      C$^{18}$O & 3--2 & 31.64 & JCMT 15-m & 23 & & 30$\pm$4 & 17 \\
      &&&&&&&\\
      $^{13}$CO & 1--0 & 5.29 & NRO 45-m & 16.5 & 299$\pm$8 & 60.6$\pm$0.4 & 1 \\
      $^{13}$CO & 2--1 & 15.88 & JCMT 15-m & 21 & 400 & 75 & 19 \\
      $^{13}$CO & 2--1 & 15.88 & SEST 15-m & 24 & 500 & 104 & 19 \\
      $^{13}$CO & 2--1 & 15.88 & JCMT 15-m & 23 & & 82$\pm$10 & 18 \\
      $^{13}$CO & 2--1 & 15.88 & IRAM 30-m & 12 & & 134.2$\pm$13.4 & 18 \\
      $^{13}$CO & 2--1 & 15.88 & HHT 10-m & 34 & 244 & 52.4$\pm$4.3 & 14 \\
      $^{13}$CO & 3--2 & 31.76 & CSO 10-m & 24 & 1000 & 210 & 19 \\
      $^{13}$CO & 3--2 & 31.76 & JCMT 15-m & 23 & & 90$\pm$13 & 18 \\
      &&&&&&&\\
      CH$_{3}$CN & 6$_{K}$--5$_{K}$ & 18.56 & NRO 45-m & 16.5 & 16$\pm$2 & 3.2$\pm$0.2 & 1 \\
      CH$_{3}$CN & 9$_{K}$--8$_{K}$ & 39.77 & IRAM 30-m & 15.0 & 7.0 & 1.20$\pm$0.30 & 2 \\
      &&&&&&&\\
      C$^{17}$O & 1--0 & 5.40 & NRO 45-m & 15.6 & 13$\pm$5 & 1.5$\pm$0.6 & 1 \\
      C$^{17}$O & 2--1 & 16.19 & IRAM 30-m & 13 & & 4.5$\pm$1.1 & 17 \\
      C$^{17}$O & 2--1 & 16.19 & IRAM 30-m & 12 & & 5.7$\pm$0.8 & 18 \\
      &&&&&&&\\
      CN & 1--0 & 5.43 & NRO 45-m & 15.2 & 422$\pm$9 & 150$\pm$1.7 & 1 \\
      CN & 2--1 & 16.32 & IRAM 30-m & 13 & & 92.9$\pm$4.2 & 20 \\
      &&&&&&&\\
      $^{12}$CO & 1--0 & 5.54 & NRO 45-m & 15.2 & 4241$\pm$71 & 869.9$\pm$2.5 & 1 \\
      $^{12}$CO & 2--1 & 16.61 & JCMT 15-m & 21 & 6100 & 1152 & 19 \\
      $^{12}$CO & 2--1 & 16.61 & SEST 15-m & 24 & 5000 & 926 & 19 \\
      $^{12}$CO & 2--1 & 16.61 & IRAM 30-m & 12 & & 1310 & 21 \\
      $^{12}$CO & 2--1 & 16.61 & JCMT 15-m & 23 & & 1062$\pm$117 & 18 \\
      $^{12}$CO & 3--2 & 33.22 & CSO 10-m & 24 & 6200 & 1194 & 19 \\
      $^{12}$CO & 3--2 & 33.22 & JCMT 15-m & 14 & & 1200 & 22 \\
      $^{12}$CO & 3--2 & 33.22 & JCMT 15-m & 23 & & 998$\pm$140 & 18 \\
      $^{12}$CO & 4--3 & 55.37 & CSO 10-m & 15 & 3300$\pm$200 & 507 & 23 \\
      $^{12}$CO & 4--3 & 55.37 & JCMT 15-m & 10.4 & & 2160 & 22 \\
      $^{12}$CO & 4--3 & 55.37 & JCMT 15-m & 22 & & 1019$\pm$120 & 24 \\
      $^{12}$CO & 6--5 & 116.26 & JCMT 15-m & 8 & 5200$\pm$900 & 861 & 25 \\
      $^{12}$CO & 6--5 & 116.26 & CSO 10-m & 10.60 & & 1394$\pm$278.8 & 26 \\
      $^{12}$CO & 7--6 & 155.01 & JCMT 15-m & 11.5 & & 1370 & 27 \\
      $^{12}$CO & 7--6 & 155.01 & CSO 10-m & 8.95 & & 810.2$\pm$162.0 & 26 \\
\end{longtable}

\begin{longtable}{lccccccc}
  \caption{IC 342.}
      \hline
      Molecule & Transition & $E\rm{u}/\it{k}$ & Telescope & $\theta_{\rm b}$ & $T_{\rm mb}$ & $\int$$T_{\rm mb}dv$ & Reference \\
       & & [K] & & [$^{\prime\prime}$] & [mK] & [K km s$^{-1}$] & \\
      \hline
      \endhead
      \hline
      \endfoot
      \hline
      \multicolumn{8}{l}{\hbox to 0pt{\parbox{130mm}{\footnotesize
       References. (1) Takano et al. in prep.; (2) Aladro et al. 2011a; (3) Nguyen-Q-Rieu et al. 1991; (4) Mart\'{i}n et al. 2009; (5) Paglione et al. 1997; (6) Nguyen-Q-Rieu et al. 1992; (7) Jackson et al. 1995; (8) Mauersberger et al. 1995; (9) Paglione et al. 1995; (10) Mauersberger \& Henkel 1989; (11) Mauersberger et al. 2003; (12) Bayet et al. 2009; (13) Eckart et al. 1990; (14) Sage et al. 1991; (15) Israel \& Baas 2003; (16) Bayet et al. 2006; (17) Henkel et al. 1988; (18) G$\ddot{\rm u}$sten et al. 1993; (19) Harris et al. 1991.
      }}}
      \endlastfoot
      CH$_{3}$CCH & 5$_{K}$--4$_{K}$ & 12.31 & NRO 45-m & 18.7 & 8$\pm$1 & 0.5$\pm$0.1 & 1 \\
      CH$_{3}$CCH & 6$_{K}$--5$_{K}$ & 17.24 & NRO 45-m & 16.5 & 8$\pm$1 & 0.6$\pm$0.1 & 1 \\
      CH$_{3}$CCH & 8$_{K}$--7$_{K}$ & 29.55 & IRAM 30-m & 18.2 & 13.7$\pm$1.5 & 0.7$\pm$0.1 & 2 \\
      &&&&&&&\\
      HNCO & 4$_{0,4}$--3$_{0,3}$ & 10.56 & NRO 45-m & 19.1 & 37$\pm$4 & 1.5$\pm$0.2 & 1 \\
      HNCO & 5$_{0,5}$--4$_{0,4}$ & 15.84 & NRO 45-m & 16.5 & 30$\pm$2 & 1.5$\pm$0.1 & 1 \\
      HNCO & 6$_{0,6}$--5$_{0,5}$ & 22.17 & IRAM 30-m & 17 & 40$\pm$10 & 2.5$\pm$0.5 & 3 \\
      HNCO & 6$_{1,6}$--5$_{1,5}$ & 65.40 & IRAM 30-m & 19 & 33.0$\pm$1.9 & 1.87$\pm$0.11 & 4 \\
      HNCO & 10$_{0,10}$--9$_{0,9}$ & 58.07 & IRAM 30-m & 11 & 8.5$\pm$1.1 & 0.31$\pm$0.06 & 4 \\
      &&&&&&&\\
      HCN & 1--0 & 4.26 & NRO 45-m & 19.1 & 183$\pm$3 & 11.6$\pm$0.1 & 1 \\
      HCN & 3--2 & 25.54 & CSO 10-m & 27 & 70$\pm$13 & 3.6$\pm$0.6 & 5 \\
      HCN & 3--2 & 25.54 & IRAM 30-m & 10 & & 7.1$\pm$1 & 6 \\
      HCN & 4--3 & 42.57 & CSO 10-m & 20 & 90$\pm$20 & 5.9 & 7 \\
      &&&&&&&\\
      HCO$^{+}$ & 1--0 & 4.28 & NRO 45-m & 19.1 & 140$\pm$3 & 8.4$\pm$0.1 & 1 \\
      HCO$^{+}$ & 3--2 & 25.70 & IRAM 30-m & 10 & & 7.2$\pm$1 & 6 \\
      &&&&&&&\\
      HC$_{3}$N & 9--8 & 19.67 & IRAM 30-m & 30.2 & 12.2$\pm$1.2 & 0.7$\pm$0.1 & 2 \\
      HC$_{3}$N & 10--9 & 24.04 & NRO 45-m & 19.1 & 17$\pm$3 & 1.1$\pm$0.2 & 1 \\
      HC$_{3}$N & 11--10 & 28.84 & NRO 45-m & 16.5 & 17$\pm$2 & 0.7$\pm$0.1 & 1 \\
      HC$_{3}$N & 12--11 & 34.09 & NRO 45-m & 16.5 & 13$\pm$1 & 0.7$\pm$0.1 & 1 \\
      HC$_{3}$N & 15--14 & 52.44 & IRAM 30-m & 18.2 & 9.8$\pm$2.0 & 0.5$\pm$0.1 & 2 \\
      HC$_{3}$N & 16--15 & 59.43 & IRAM 30-m & 16 & 8$\pm$3 & 0.3$\pm$0.1 & 8 \\
      HC$_{3}$N & 16--15 & 59.43 & IRAM 30-m & 16.9 & 6.7$\pm$1.8 & 0.4$\pm$0.1 & 2 \\
      HC$_{3}$N & 17--16 & 66.86 & IRAM 30-m & 16.0 & 6.6$\pm$1.5 & 0.7$\pm$0.1 & 2 \\
      HC$_{3}$N & 24--23 & 131.09 & IRAM 30-m & 11.5 & 8.4$\pm$2.8 & 0.7$\pm$0.2 & 2 \\
      &&&&&&&\\
      C$^{34}$S & 2--1 & 6.95 & NRO 45-m & 19 & 10$\pm$1 & 0.4$\pm$0.1 & 1 \\
      C$^{34}$S & 3--2 & 13.89 & IRAM 30-m & 16 & 16$\pm$2 & 0.74$\pm$0.12 & 8 \\
      C$^{34}$S & 3--2 & 13.89 & IRAM 30-m & 17 & 8.4$\pm$1.5 & 0.49$\pm$0.12 & 4 \\
      &&&&&&&\\
      CH$_{3}$OH & 2$_{K}$--1$_{K}$ & 6.97 & NRO 45-m & 19 & 53$\pm$1 & 2.8$\pm$0.1 & 1 \\
      &&&&&&&\\
      CS & 1--0 & 2.35 & NRO 45-m & 36 & 50$\pm$12 & 2.5$\pm$0.4 & 9 \\
      CS & 2--1 & 7.06 & NRO 45-m & 19 & 96$\pm$1 & 5.1$\pm$0.1 & 1 \\
      CS & 3--2 & 14.12 & IRAM 30-m & 16.8 & 88.5$\pm$3.0 & 4.9$\pm$0.1 & 2 \\
      CS & 5--4 & 35.30 & IRAM 30-m & 11 & 35 & 0.76$\pm$0.22 & 10 \\
      CS & 5--4 & 35.30 & HHT 10-m & 32 & 8.5 & 0.78$\pm$0.17 & 11 \\
      CS & 5--4 & 35.30 & IRAM 30-m & 10 & 40.3$\pm$1.8 & 2.15$\pm$0.11 & 4 \\
      CS & 5--4 & 35.30 & IRAM 30-m & 10.1 & 37.6$\pm$2.9 & 3.4$\pm$0.1 & 2 \\
      CS & 7--6 & 65.88 & JCMT 15-m & 14.0 & 12.3$\pm$8.5 & 1.0$\pm$0.4 & 12 \\
      &&&&&&&\\
      SO & 3$_{2}$--2$_{1}$ & 9.23 & NRO 45-m & 16.5 & 13$\pm$1 & 0.9$\pm$0.1 & 1 \\
      SO & 2$_{3}$--1$_{2}$ & 21.07 & NRO 45-m & 16.5 & & $<$0.07 & 1 \\
      SO & 4$_{5}$--4$_{4}$ & 24.45 & NRO 45-m & 16.5 & & $<$0.07 & 1 \\
      &&&&&&&\\
      C$^{18}$O & 1--0 & 5.27 & NRO 45-m & 16.5 & 77$\pm$2 & 4.2$\pm$0.1 & 1 \\
      C$^{18}$O & 2--1 & 15.82 & IRAM 30-m & 14 & & 8.4$\pm$1.7 & 13 \\
      C$^{18}$O & 2--1 & 15.82 & IRAM 30-m & 13 & & 4.2$\pm$0.6 & 14 \\
      C$^{18}$O & 2--1 & 15.82 & IRAM 30-m & 11 & 164.1$\pm$1.1 & 8.10$\pm$0.10 & 4 \\
      &&&&&&&\\
      $^{13}$CO & 1--0 & 5.29 & NRO 45-m & 16.5 & 294$\pm$3 & 16.2$\pm$0.1 & 1 \\
      $^{13}$CO & 2--1 & 15.88 & IRAM 30-m & 14 & & 29.7$\pm$1.1 & 13 \\
      $^{13}$CO & 2--1 & 15.88 & JCMT 15-m & 21 & 476 & 24.0$\pm$3 & 15 \\
      $^{13}$CO & 2--1 & 15.88 & HHT 10-m & 34 & 253 & 17.2$\pm$1.1 & 11 \\
      $^{13}$CO & 3--2 & 31.76 & JCMT 15-m & 14 & 311 & 17.1$\pm$2 & 15 \\
      $^{13}$CO & 3--2 & 31.76 & CSO 10-m & 21.90 & & 19.9$\pm$2.2 & 16 \\
      &&&&&&&\\
      CH$_{3}$CN & 6$_{K}$--5$_{K}$ & 18.56 & NRO 45-m & 16.5 & 8$\pm$2 & 0.6$\pm$0.1 & 1 \\
      &&&&&&&\\
      C$^{17}$O & 1--0 & 5.40 & NRO 45-m & 15.6 & 9$\pm$3 & 0.7$\pm$0.2 & 1 \\
      C$^{17}$O & 2--1 & 16.19 & IRAM 30-m & 13 & & 0.76$\pm$0.19 & 14 \\
      &&&&&&&\\
      CN & 1--0 & 5.43 & NRO 45-m & 15.2 & 149$\pm$4 & 20.4$\pm$0.4 & 1 \\
      CN & 2--1 & 16.32 & IRAM 30-m & 13 & & 5.2$\pm$0.2 & 17 \\
      &&&&&&&\\
      $^{12}$CO & 1--0 & 5.54 & NRO 45-m & 15.2 & 2671$\pm$18 & 160.0$\pm$0.4 & 1 \\
      $^{12}$CO & 2--1 & 16.61 & IRAM 30-m & 14 & & 324.3$\pm$3.2 & 13 \\
      $^{12}$CO & 2--1 & 16.61 & JCMT 15-m & 21 & 3090 & 172$\pm$19 & 15 \\
      $^{12}$CO & 3--2 & 33.22 & JCMT 15-m & 14 & 3140 & 186$\pm$23 & 15 \\
      $^{12}$CO & 3--2 & 33.22 & CSO 10-m & 21.9 & & 109.8$\pm$2.5 & 16 \\
      $^{12}$CO & 4--3 & 55.37 & CSO 10-m & 15.1 & 2200$\pm$300 & 88 & 18 \\
      $^{12}$CO & 4--3 & 55.37 & JCMT 15-m & 11 & 4020 & 209$\pm$21 & 15 \\
      $^{12}$CO & 4--3 & 55.37 & CSO 10-m & 14.55 & & 110.8$\pm$12.6 & 16 \\
      $^{12}$CO & 6--5 & 116.26 & JCMT 15-m & 8 & 1200$\pm$400 & 98 & 19 \\
      $^{12}$CO & 6--5 & 116.26 & CSO 10-m & 10.60 & & 54.6$\pm$3.3 & 16 \\
      $^{12}$CO & 7--6 & 155.01 & CSO 10-m & 8.95 & & 30.2$\pm$2.8 & 16 \\
\end{longtable}

\end{document}